\documentclass[sigconf]{acmart}

\AtBeginDocument{%
  \providecommand\BibTeX{{%
    \normalfont B\kern-0.5em{\scshape i\kern-0.25em b}\kern-0.8em\TeX}}}

\copyrightyear{2025}
\acmYear{2025}
\setcopyright{rightsretained}
\acmConference[CHI EA '25]{Extended Abstracts of the CHI Conference on Human Factors in Computing Systems}{April 26-May 1, 2025}{Yokohama, Japan}
\acmBooktitle{Extended Abstracts of the CHI Conference on Human Factors in Computing Systems (CHI EA '25), April 26-May 1, 2025, Yokohama, Japan}
\acmDOI{10.1145/3706599.3719858}
\acmISBN{979-8-4007-1395-8/2025/04}

\begin{document}


\title[Improving UX with FAICO]{Improving User Experience with FAICO: Towards a Framework for AI Communication in Human-AI Co-Creativity}

\author{Jeba Rezwana}
\email{jrezwana@towson.edu}
\orcid{0000-0003-1824-249X}
\affiliation{%
  \institution{Towson University}
  \city{Towson}
  \country{USA}
}

\author{Corey Ford}
\email{c.ford@arts.ac.uk}
\orcid{0000-0002-6895-2441}
\affiliation{%
  \institution{University of the Arts London}
  \city{London}
  \country{United Kingdom}
}

\renewcommand{\shortauthors}{}

\begin{abstract}

How AI communicates with humans is crucial for effective human-AI co-creation. However, many existing co-creative AI tools cannot communicate effectively, limiting their potential as collaborators. This paper introduces our initial design of a Framework for designing AI Communication (FAICO) for co-creative AI based on a systematic review of 107 full-length papers. FAICO presents key aspects of AI communication and their impacts on user experience to guide designing effective AI communication. We then show actionable ways to translate our framework into two practical tools: design cards for designers and a configuration tool for users. The design cards enable designers to consider AI communication strategies that cater to a diverse range of users in co-creative contexts, while the configuration tool empowers users to customize AI communication based on their needs and creative workflows. This paper contributes new insights within the literature on Human-AI Co-Creativity and Human-Computer Interaction, focusing on designing AI communication to enhance user experience.
\end{abstract}




\maketitle


\section{Introduction}

Human-AI co-creativity involves humans and AI collaborating in a creative process as partners to produce creative artifacts, ideas or performances \cite{davis2013human}. Co-creativity research suggests \cite{liapis2014computational} that when creativity emerges from human-AI interaction, it can surpass contributors' original creativity and intentions as novel ideas arise in the process. With the emergence of popular Generative AI (GenAI) systems with co-creative abilities such as ChatGPT \cite{openaiChatGPTOptimizing}, DALL-E \cite{openaiDALLE} and Midjourney \cite{midjourney}, human-AI co-creativity is making its way into mainstream life. We suggest that the next frontier of co-creative AI needs good \textbf{collaborative skills} in addition to algorithmic competence. However, designing co-creative AI has \textbf{many challenges} due to the open-ended nature of the interaction between humans and AI in creative contexts \cite{davis2016empirically, kantosalo2014isolation}. For example, co-creative AI must be able to adapt to spontaneous interaction styles and allow creative products to develop dynamically across different parts of the creative process. 

An essential component in human-AI co-creativity is \emph{communication} for the co-regulation of collaborators \cite{bown2020speculative}. In this paper, we define \textbf{\emph{AI Communication}} as where an AI interacts with humans purposefully and directly to convey information \cite{rezwana2022designing}. Our focus is specifically on direct communication, excluding indirect communication through sensemaking that occurs through creative contributions. For instance, in a co-creative drawing scenario, AI might provide feedback on users' contributions or proactively suggest design improvements, such as saying, ``The house you drew is a little out of proportion. Would you like to fix it, or should I adjust it for you?'' The user can choose to incorporate or disregard this suggestion, which distinguishes AI communication from AI's contribution to the creative process, such as drawing a tree next to the user's house. Our definition of AI communication also overlaps with the field explainable AI \cite{DARPA, gunning_2016}, in particular considering how AI communication can provide additional explanations in co-creative AI contexts \cite{xaixarts_workshop}. For example, we consider moments where an AI might communicate to people at the moment whilst they interact with the creative product \cite{xaixarts_workshop_2}.

For effective co-creation, \citet{mamykina2002collaborative} suggests that co-creative AI should provide feedback and critique on contributions similar to how human collaborators do in teamwork. When technology effectively communicates with people 
 HCI research has shown that they are perceived as independent \emph{social actors} \cite{nass1994computers}, improving user perception of AI \cite{rezwana2022understanding}. Therefore, \emph{AI Communication} can be utilized for co-creative AI to be perceived as an equal partner in co-creation \cite{mcmillan2002measures,rezwanacofi} and to improve user experience \cite{rezwana2022understanding}. However, it is challenging to design \emph{AI Communication} due to the complex and dynamic nature of the human-AI partnership \cite{kantosalo2014isolation}, which requires conveying intricate processes and suggestions across a range of different types of users \cite{weld2018intelligible}. There is a gap in HCI research of tools, such as design frameworks, for effective \emph{AI Communication} in human-AI co-creativity. Motivated by this research gap, this paper identifies the aspects of AI communication and how these aspects impact user experience through a systematic literature review. Below are the \textbf{contributions} of this research:

\begin{itemize}
    \item A Framework for AI Communication (FAICO) as the first step towards designing and interpreting \emph{AI Communication} in co-creative contexts based on a systematic literature review. FAICO identifies key aspects of \emph{AI Communication} that should be considered for effective co-creation and their influence on user experience.
    \item Translating FAICO into practical tools such as design cards for AI practitioners and a configuration tool for users, thereby fostering
    the advancement of human-centered co-creative AI systems. 
    
\end{itemize}

The paper is organized as follows. We first provide background on communication in human-AI co-creativity, followed by the methodology of our literature review. We then introduce FAICO in section 4 and move to translating FAICO into design cards and configuration tools in Section 5. Lastly, we discuss the implications and future work.








\section{Communication in Human-AI Co-Creativity}

AI ability alone does not ensure a positive user experience with the AI \cite{louie2020novice}, especially where interaction is essential between humans and AI \cite{wegner1997interaction}. \textbf{Co-creative AI} systems involve humans and AI collaborating as partners \cite{davis2013human} with creativity emerging that surpasses either party's individual creativity \cite{liapis2014computational}. In co-creative settings, effective interaction is crucial \cite{kantosalo2020modalities,bown2015player}, e.g. Karimi et al. found an association between human-AI interaction and creativity that emerges from the co-creation \cite{karimi2020creative}. However, there is a scarcity of research focused on the interaction between humans and AI in the field of co-creativity \cite{rezwana2022designing}. To address this gap, a few interaction frameworks have been proposed \cite{Guzdial2019, kantosalo2020modalities,rezwanacofi}. Notably, \citet{rezwana2022designing} introduced the COFI framework for designing and evaluating interaction design in human-AI co-creativity, highlighting communication between humans and AI as a key interaction design component. More recently, the User-Centered Framework for Human-AI Co-Creativity (UCCC) has been proposed, identifying key dimensions for modulating control between users and AI \cite{moruzzi2024user}. However, none of these frameworks specifically explore the dimensions of AI communication and their impact on user experience.  In this paper, we extend the existing frameworks by focusing on the critical role of AI communication, identifying its key dimensions, and examining how these dimensions influence user experience.

Exploring communication with computer systems has a long history, dating back to when Alan Turing \cite{french2000turing} proposed that a machine that communicates indistinguishably to humans might be considered intelligent. For example, a significant challenge in human-AI collaboration is the development of common ground for communication between humans and machines \cite{dafoe2021cooperative}. We take the view that communication is an essential component in human-AI co-creativity for the co-regulation of the collaborators \cite{liang2019implicit, bown2020speculative}. In human-AI collaboration, \emph{AI Communication} helps the AI agent make decisions in a creative process \cite{bown2020speculative} and supports many aspects of the \textbf{user experience} \cite{mcmillan2002measures}. For example, AI Communication in co-creation can improve user engagement, collaborative experience and user perception of a co-creative AI \cite{rezwana2022understanding, bryan2012identifying}. \textit{AI Communication} can also improve social presence and interpersonal trust \cite{bente2004social}. 

Designing human-like communication has been described as an underestimated challenge of machine intelligence \cite{healey2021human}. HCI research shows that how users talk in a human-AI conversation is similar to human-human conversation \cite{dev2020user}. While it is vital to use human communication and its ideas as a starting point for designing \textit{AI Communication}, it should not impose permanent restrictions on \textit{AI Communication} \cite{guzman2020artificial}. For instance, communication between artists is often based on intuition and non-verbal interaction \cite{healey2005inter}, rather than direct explanation-based discussions which are more traditionally seen in non-creative contexts \cite{xaixarts_workshop}. Identifying the appropriate communication for co-creative AI is thus key to its user experience. For example, elements of communication such as voice elicit human-like responses toward devices \cite{nass2005wired}, which may or may not always be appropriate depending on context. 
In this paper, we introduce our first steps in creating a framework that could be helpful to guide the design of \textit{AI Communication} in co-creation.





\section{Framework Development Methodology: A Systematic Literature Review}

In this section, we document the systematic literature review approach we followed as a methodology to identify the key aspects of \textit{AI Communication} and their influence on user experience. 

To identify essential aspects of \textit{AI Communication}, we used the ACM Digital Library to connect to conferences where HCI, AI and creativity research is typically published and to build on reviews by others surveying creativity support \cite{frich2019}. We considered documents published from 1995 until 2024 to consider recent articles. We narrowed our search to full-length papers (e.g. not tutorials or posters) to ensure we considered high-quality peer-reviewed studies. We also only considered papers in English that could be understood by both authors.

We used keywords aligned with our research goal for the framework to search for relevant academic publications. We (the paper authors) chose these keywords following several weekly discussions where we reflected upon our own experiences of AI communication as co-creative AI researchers. The first two keywords related directly to communication:  ``Aspects of Communication'' (561 results) and ``Factors of Communication'' (41 results). The next three related to interaction and communication between AI and Humans: ``Human-AI Communication'' (47 results), ``AI Communication'' (117 results)  and ``Human-AI Interaction'' (279 results). The final four related to when the AI might communicate with humans: ``Proactive Communication'' (90 results), ``Reactive Communication'' (29 results), ``Proactive AI'' (19 results) and ``Reactive AI'' (13 results). From reviewing the abstracts of these papers, we collected 132 papers. 

\begin{figure*}[h]
    \centering
    \includegraphics[width=1\linewidth]{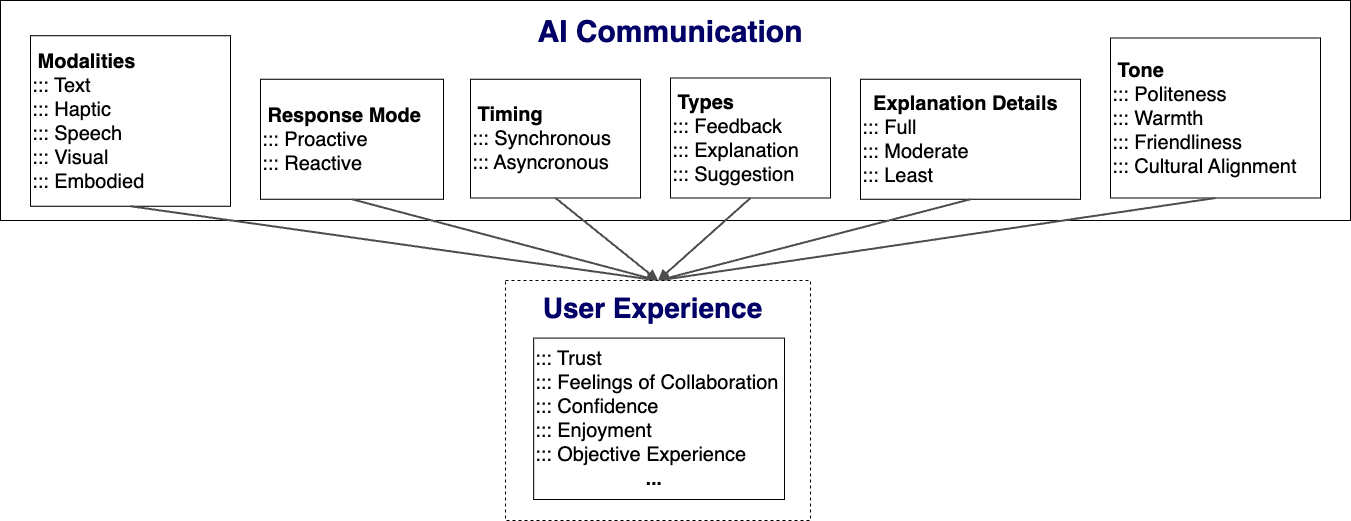}
    
    \caption{The Framework for AI Communication in Co-Creative Contexts\newline{}}
    \vspace{-0.5cm}
    \label{framework}
    \Description{The Framework for AI Communication in Co-Creative Contexts (FAICO)}
\end{figure*}

Next, we read the 132 papers, narrowing down the corpus to \textbf{107 papers} with relevance to AI Communication. We summarized the discussion on how co-creative AI should communicate information to users and used affinity diagramming on these descriptions \cite{harboe2015affinity} to organize the summaries into related groups. Through iterative discussion, we created the first draft of our framework. The framework described below is primarily based on our literature review while also integrating relevant insights from additional works outside the scope of the review. The formation and description of FAICO follow a methodology \textbf{similar} to that used in other interaction and user-centered frameworks for human-AI co-creation, such as COFI by Rezwana and Maher \cite{rezwana2022designing} and UCCC by Moruzzi and Margarido \cite{moruzzi2024user}. 





\section{Framework For AI Communication (FAICO)}
Figure \ref{framework} shows the Framework for AI Communication (FAICO) in the context of human-AI co-creation, introducing \textbf{critical aspects} to consider for AI communication and their influence on the \textbf{user experience}. The dotted line for user experience (Fig \ref{framework}) indicates that it is not directly a part of AI Communication but rather comprises aspects influenced by it. Below we discuss each aspect of FAICO and their influence and connections to user experience based on our literature review and related work.

\subsection{Modalities} 

According to FAICO, \textit{Modalities} refers to distinct channels through which AI agents communicate with humans, which is supported by HCI research \cite{karray2008human}. Rezwana and Maher outline the classification of modalities for AI to human communication in human-AI co-creation, encompassing text, speech, visuals, haptic, and embodied communication \cite{rezwana2022designing}. In our framework, we adopt their classification of modalities for AI Communication.

\subsubsection{Connection to User Experience}

The literature demonstrates that communication modalities influence user experience. Text-based dialogue plays a key role in helping users perceive the AI's personality \cite{mairesse2007using} and enhances user satisfaction and enjoyment \cite{oh2018lead}. Speech-based communication has been shown to improve perceived communication quality and trust, often performing better than visual cues in enhancing user experience \cite{zhang2024verbal}. Verbal and non-verbal cues allow AI agents to convey their states effectively, which contributes to improved feelings of collaboration and fosters effective human-AI partnerships \cite{anzabi2023effect,bryan2012identifying, sonlu2021conversational}. Haptics and visual cues also play an important role, enhancing users’ understanding of emotional qualities and fostering a sense of connection \cite{wang_haptic_2023, price_touch_2022, rodrigues_emoji_2022, nanayakkarawasam_icon_2023}. Embodied communication improves the feelings of collaboration with perceived presence compared to text-based communication \cite{lim2024artificial}. Furthermore, multimodal communication, such as combining visuals with verbal interactions, has been found to improve user experience \cite{samek2023cosy}. For instance, the integration of voice, embodied communication, dialogue, and facial expressions provides users with a more accurate perception of AI, improving feelings of collaboration and social presence \cite{sonlu2021conversational, munnukka_2022}.

The influence of modalities in AI Communication varies between user demographics such as gender \cite{huang2021women}, age \cite{razavi2022discourse}, neurodiversity \cite{samek2023cosy}, as well as the context in which multimodality is introduced \cite{esau2023foggy}. 
For example, \citet{vossen2009social} found a gender-specific response to embodied communication. To improve the user experience for a diverse population, \textbf{multimodal interactions} significantly enhance the performance of AI compared to unimodal interactions \cite{samek2023cosy, park2021toddler}.

\subsection{Response Mode}

\textit{Response Mode} refers to the approach used by an AI agent to initiate or respond to communication during a co-creation. AI can be either \emph{proactive} or \emph{reactive} in terms of how it responds. Proactive AI systems ``do not wait for user input but instead actively initiate communication'' \cite{van2021proactive}. On the other hand, when an AI is reactive, it communicates with users when prompted, for example, when clicking a button \cite{meurisch2020exploring}.

\subsubsection{Connection to User Experience}
Response mode influences user experience in different ways. Proactive AI, while generally favored in social interactions over reactive AI \cite{meurisch2020exploring}, has been perceived as annoying in several studies \cite{van2021proactive}, particularly in contexts where it interrupts users' immersion in creative activities \cite{csikszentmihalyi_flow_1990} where reactive AI might be preferable. Reactive AI can improve task performance over time and improve perceived usefulness \cite{kim2021ai, berry2006can}. However, proactive AI can improve perceived trust \cite{zhang2023investigating, kraus2022kurt,jain2023co} and encourage users to be more critical of AI output, which we found connected with human-centered AI guidelines on responsible design \cite{ozmen2023sixchallenges}.

The demographics of users and contexts also affected the different types of timing participants wanted. \citet{pang2013technology} found that while most mental health app users preferred reactive AI, researchers like doctoral students were more open to proactive AI. Similarly, \citet{meurisch2020exploring} demonstrated that factors such as age, country, gender, and personality traits impact user preferences, with individuals with higher openness favoring proactive AI, while those lower in extraversion preferred reactive communication. \citet{luria2022letters} further showed that parents' desired level of AI proactivity varied depending on the situation, emphasizing the importance of context.

\subsection{Timing} 

\textit{Timing} refers to \emph{when} an AI system communicates with the user, describing whether interactions occur simultaneously or at different times. Timing has been indicated as a crucial aspect of communication in human-AI collaboration \cite{zhou2024impact, fan2022human}. AI Communication can happen either in a \emph{synchronous} (whilst co-creating) or \emph{asynchronous} (outside of working times such as notifications) manner \cite{fan2022human}. 

\subsubsection{Connection to User Experience}
Timing affects the user experience in many different ways. Synchronous communication is beneficial when immediate feedback and real-time collaboration are necessary \cite{lim2017analysis}. It's particularly useful in improvisational co-creation, such as the studio stage of choreography creation, where AI can engage with users in real-time \cite{liu2024dancegen}. Situations requiring a spontaneous and immediate exchange of ideas and feedback from AI can enhance the co-creation process and successful task completion \cite{el2022sand}. Asynchronous communication can be useful when users need time for reflection and deeper consideration of AI suggestions \cite{striner2022co,berry2024reactive}. Creative activities that benefit from thoughtful consideration, such as reviewing creative writing or art and producing reflection, asynchronous communication from the AI, could be useful there \cite{xia2023review,liu2024dancegen}.

Timing influences different user demographics and creative stages differently. Experienced users often prefer asynchronous communication from AI \cite{wu2021ai}, likely because it allows them to maintain an uninterrupted creative flow. Different stages of the creative process may require different timing of AI communication \cite{wu2021ai}. For example, the ideation stage can benefit from synchronous communication, but the execution stage, where AI works on time-consuming tasks, can use asynchronous communication \cite{hwang2022too, lin2022creative}.

\subsection{Communication Type} 

Communication type refers to the type of AI Communication with which AI systems convey information to users based on distinct purposes. The type of communication as one of the major dimensions in AI-mediated communication \cite{guzman2020artificial,hancock2020ai}. According to FAC, types of AI Communication include: \emph{explanation}, \emph{suggestion}, and \emph{feedback}. \emph{Explanation} can be defined as representations of underlying causes that led to a system’s output and which reflect decision-making processes. \emph{Feedback} refers to communication in the form of critique, analysis, or assessment of a particular contribution, idea, or artifact. 
\emph{Suggestion} refers to recommendations or alternatives to the user contributions during collaborative processes that align with the user's creative objectives.  


\subsubsection{Connection to User Experience}
Research shows that communication type influences various phases of co-creation. Example-based suggestions are effective in the early stages, fostering ideation, while rule-based suggestions are better suited for promoting user understanding \cite{yeh2022guide}. Divergent suggestions are particularly useful during the initial phases to encourage creativity, whereas convergent suggestions are more effective later, helping to refine the co-created product \cite{karimi2020creative}. AI explanations also play a crucial role in shaping user trust. Clear reasoning enhances trustworthiness \cite{mehrotra2024systematic}, and social transparency, such as making social contexts visible, improves both trust and decision-making \cite{ehsan2021expanding}. However, while providing reasoning increases trust, disclosing uncertainty can diminish it \cite{vossing2022designing}. Excessive trust may lead to overreliance on AI, reducing human accuracy \cite{nguyen2018believe}, a problem further amplified by human-like behaviors, such as hesitations in chatbot responses \cite{zhou2024beyond}. Moreover, users tend to value explanations of overall decision-making more than detailed explanations of individual actions, particularly in team settings \cite{zhang2021ideal}.

Users' responses to AI suggestions are shaped by demographics and task-related factors, including self-confidence, confidence and alignment with the suggestion, and pre-existing beliefs about human versus AI performance \cite{vodrahalli2022humans}. Tailoring AI communication to align with users' cultural backgrounds further enhances interaction effectiveness \cite{zhao2024tailoring}.

\subsection{Explanation Details}

\emph{Explanation details} refers to the level of detail that users are given by an AI in co-creation. AI Communication can range from giving \textit{full}, to \textit{moderate}, to the \textit{least} amount of details supported by \citet{moruzzi2024user}.

\subsubsection{Connection to User Experience}
How much explanation detail is useful seems to vary depending on the co-creative context. There is a balance to be struck between detailed explanations and more ambiguous explanations that offer a moderate understanding of how an AI works whilst still providing an opportunity for serendipity or surprise \cite{xaixarts_workshop}. For example, in AI music making, whilst AI ambiguity can lead to moments of frustration, it can also be appropriated by musicians \cite{using_incongrous}. With an AI for co-creative drawing, \citet{oh2018lead} found that whilst users were more content with drawings based on detailed instructions from a co-creative AI, they also enjoyed its unpredictable nature. Full explanations are often beneficial during ideation stages, helping users understand the AI's intentions \cite{mccormack_silent_way,using_incongrous,xaixarts_workshop}, whereas less detailed explanations are preferable for in-the-moment interactions, such as reflecting on the AI's output, where additional context may not be necessary \cite{ford_reflection_across}.

The appropriate level of explanation also depends on user demographics and the stage of the creative process. Explanations should be tailored to the user's AI literacy and expertise \cite{moruzzi2024user,xaixarts_workshop_2}. For instance, experienced users may prefer full explanations to access the detailed sequence of actions performed by the AI \cite{zhu2018explainable}, while non-experts may benefit more from moderate or minimal explanations to avoid overwhelming complexity \cite{moruzzi2024user}.

\begin{figure*}[h]
    \centering
    \begin{minipage}[c]{0.57\linewidth}
        \centering
        \includegraphics[width=0.98\linewidth]{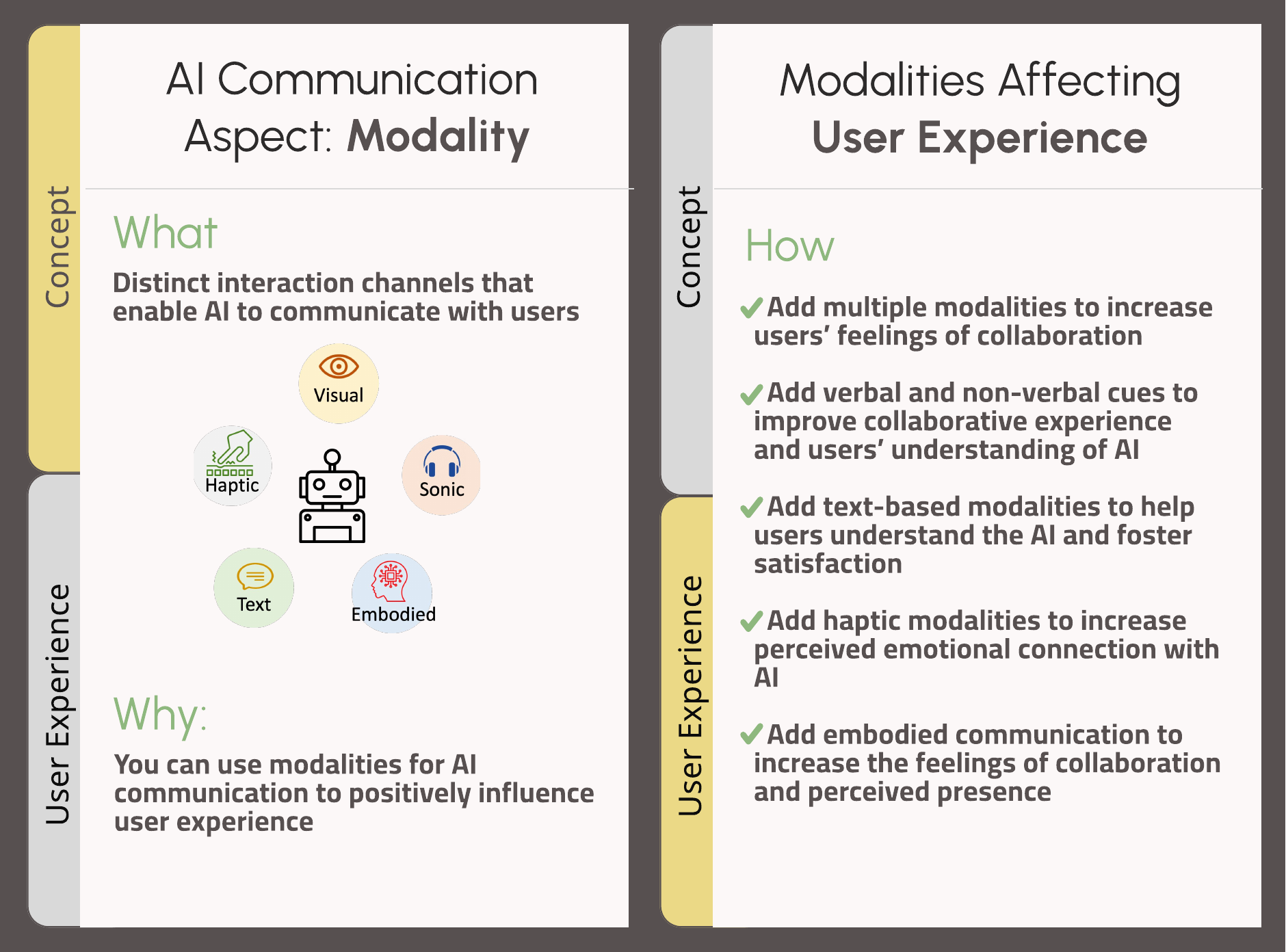}
        \caption{Design Cards}
        \Description{Design cards for the designers as a use-case of FAICO}
        \label{Design_card}
    \end{minipage}
    \hspace{0.01\linewidth}
    \begin{minipage}[c]{0.41\linewidth}
        \centering
        \includegraphics[width=\linewidth]{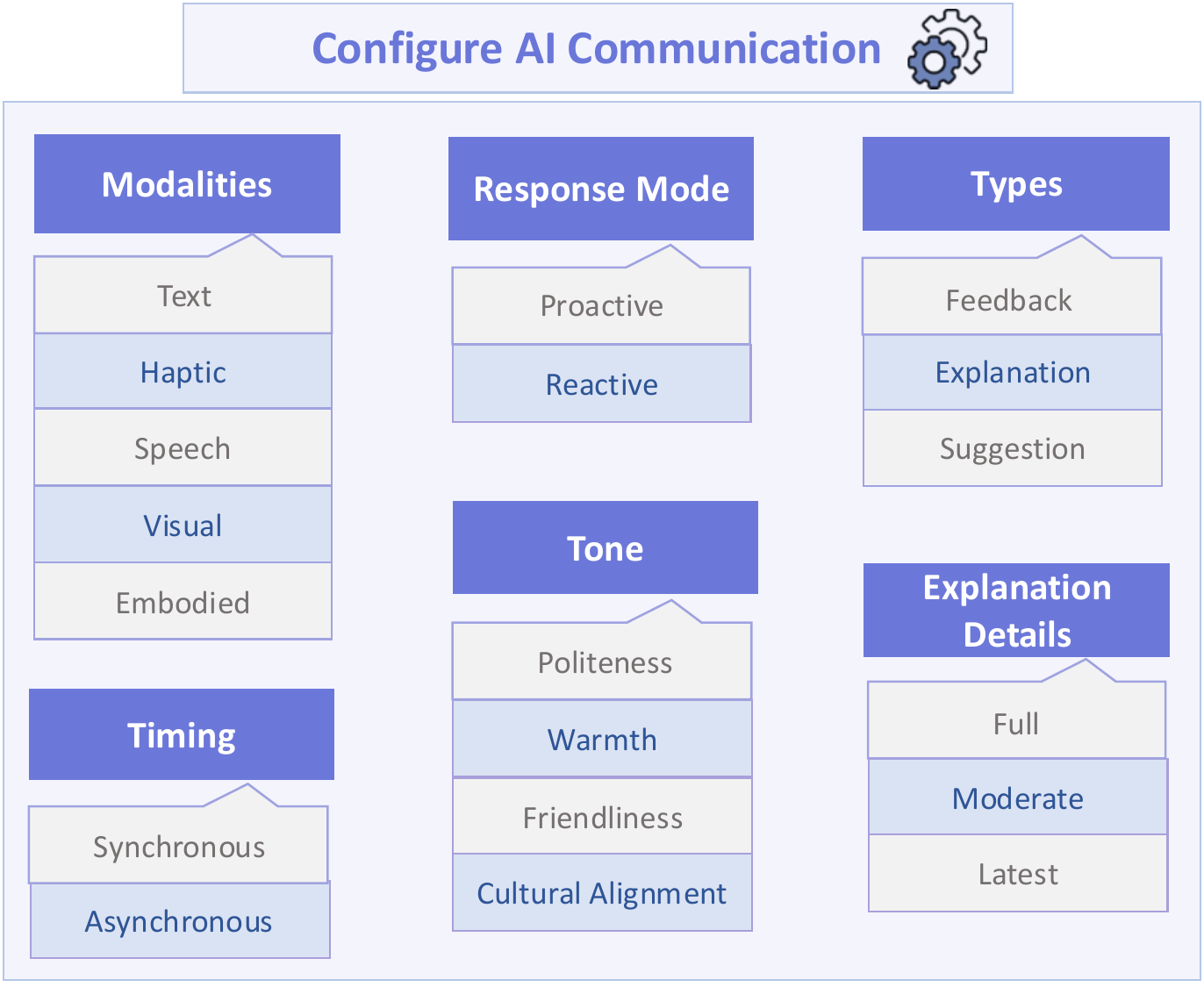}
        \caption{Configuration Tool for Users}
        \Description{Configuration tool for users as a use case of FAICO}
        \label{customization}
    \end{minipage}
\end{figure*}

\subsection{Tone} 
\textit{Tone} refers to qualities of AI Communication that determine whether the expression of the communication conveys a positive emotional intent or not. Tone has been suggested as an aspect that can affect human-AI team performance \cite{mallick2024pursuit, leong2023exploratory}. In our literature search, we found four influences on the perceived tone of AI Communication - \emph{politeness} (use of considerate, respectful language and behavior following social norms), \emph{warmth} (being approachable and affable in interactions), \emph{friendliness} (ability to foster congenial relationships) and \emph{cultural alignment} (following practices and norms matching the expectations of those involved). 

\vspace{-0.2cm}
\subsubsection{Connection to User Experience}
The tone of AI communication significantly impacts user experience. \citet{colley2021} found that when virtual AI avatars in an autonomous vehicle simulation communicated acknowledgment of a user’s polite gestures (e.g., waving thank you), the positive perception of the AI increased, demonstrating how respectful behaviors and adherence to cultural norms enhance user experience. Warmth and friendliness also improve user satisfaction and perceptions of AI capability. For instance, \citet{gilad2021warmth} showed that increasing an AI’s warmth positively influenced user satisfaction, whereas \citet{poeller2023suspecting} found that overly positive messages were sometimes met with skepticism in video game contexts, emphasizing the importance of context in determining the appropriateness of tone. Similarly, AI that responds in a human-like manner with supportive and friendly communication fosters a more engaging and satisfying user experience compared to interactions that lack these qualities \cite{d2013autotutor}.

\citet{mascarenhas2009cultural} demonstrated that variations in AI agents' rituals based on cultural behaviors had noticeable effects on users. Likewise, users preferred conversational AI aligned with their cultural norms and were adept at identifying cultural differences in dialogue \cite{endrass2009culture}. Aligning the tone of AI communication with demographic factors such as age, gender, and cultural background can lead to more effective and satisfying interactions \cite{sicilia2024humbel}.

\section{Framework Applications}
This section presents two practical applications of FAICO: Design cards for designers and configuration tools for users.

\subsection{Design Cards}


We demonstrate an actionable approach to using FAICO for designers by translating it into design cards \textbf{(Figure \ref{Design_card})}. Frameworks can be used to guide the design and scaffold of design cards; we offer the example for FAICO here to directly showcase this potential use case. These design cards could be used by AI practitioners to consider how to include effective AI communication in their co-creative AI to improve user experience.
Design cards have long been recognized as valuable tools in HCI and design work, appreciated for being ``simple, tangible, and easy to manipulate'' \cite{lucero2016designing} and an ``approachable way to introduce information and sources of inspiration as part of the design process'' \cite{wolfel2013method}. Design cards can support designers across various stages of design, including ideation, prototyping, and implementation \cite{hsieh2023cards}.

We suggest that each design card highlights a specific key component of AI communication from FAICO, explains the aspect, why it is critical to consider in designing AI communication, and how it could influence the user experience. Figure \ref{Design_card} presents a prototypical design card as a proof of concept for developing design cards from FAICO, illustrating how the cards could appear digitally. The card features two tabs: a \textit{Concept} tab describing the corresponding component of the AI Communication framework and a \textit{User Experience} tab detailing how that component impacts different aspects of user experience. In a physical format, the front of each card will include the \textit{Concept} tab, while the back will feature the \textit{User Experience} tab.

Designers can use the design cards for direction and inspiration, helping them enhance user experience through AI communication within their own projects. These cards support designers in exploring the design space of AI communication in co-creation contexts. Additionally, they provide insights into how each component of AI communication impacts different aspects of user experience, offering actionable suggestions for directly improving user experience outcomes.

\subsection{Configuration Tool} 
This section presents an approach to transforming FAICO into a configuration tool \textbf{(Figure \ref{customization})} for end-users. With the configuration tool, users can customize and personalize AI communication in their own human-AI collaboration context. The tool can be applied to various co-creative systems, such as design, writing, music composition, or art generation, enabling users to tailor AI communication to their specific needs and preferences. As highlighted in the framework section (section 4), user demographics significantly influence preferences for different aspects of AI communication. Research suggests that users who design their own products using customization toolkits perceived higher value in their creations compared to standard offerings \cite{franke2004value}. Allowing users to experience different interface configurations results in interfaces that enhance user engagement and satisfaction \cite{hui2008toward}.

The configuration tool enables users to adjust key communication parameters from FAICO, including modalities, timing, tone, response modes, and communication types, based on their unique requirements and creative objectives. For example, in a co-creative design tool, users could configure the AI to provide synchronous verbal suggestions during brainstorming sessions while opting for asynchronous text-based feedback for refining ideas at a later stage. This flexibility ensures that the AI's communication aligns with the user's workflows and supports their engagement.


\section{Discussion and Conclusions}
Despite the rapid advancement of generative AI in creative domains, many co-creative AI systems \textbf{lack effective AI communication} \cite{rezwana2022designing}, which limits the user experience \cite{bryan2012identifying, rezwana2022understanding}. Designing effective human-AI interactions requires identifying key challenges, understanding the unique complexities of AI design, and generating insights to guide future research \cite{yang2020re}. This paper presents a novel design framework for designing effective AI Communication (FAICO) to improve user experience in human-AI co-creation. Through a systematic literature review, we designed FAICO to support people in designing and configuring human-centered AI Communication by considering different aspects of AI Communication and their influence on user experience. We then illustrate how FAICO can be practically used through the creation of design cards to help designers consider AI communication in their AI systems. Additionally, we show how FAICO could be utilized as a configuration tool for co-creative AI systems, which users could use to customize and adjust AI communication preferences to align with their own goals. 

FAICO can also be used as an evaluation tool to systematically \textbf{interpret and assess} the design of AI communication in existing co-creative AI systems. Using FAICO, researchers and practitioners could identify trends, strengths, and gaps in communication design, providing valuable insights into how systems approach communication and highlighting areas for improvement to enhance user experience and future research directions. For instance, FAICO can reveal whether a system lacks proactive communication capabilities or uses only limited tone and timing preferences. Additionally, FAICO can serve as a benchmarking tool, allowing comparisons between co-creative AI systems' AI communication.


%



\subsection{Limitations \& Future Work}


FAICO is clearly in the preliminary stage and has room for refinement or extension. Additional aspects of AI communication could be incorporated to address more nuanced design requirements or to cater to emerging, niche co-creative AI domains. FAICO is also based on a systematic literature review, limited to papers from the ACM Digital library, and currently lacks empirical validation; we plan to conduct user studies with designers and users of co-creative AI to evaluate FAICO's effectiveness across diverse co-creative domains. These studies will also explore how AI communication should adapt to various co-creative contexts. Insights from the literature review suggest factors, such as the creative phase, user expertise, and domain-specific practices, influence the optimal design strategies for AI communication. Empirical studies will help better understand the influence of these factors and refine FAICO accordingly.

In addition to validating the framework itself, we plan to evaluate the utility of the design cards presented. By conducting user studies with designers using these cards to develop AI communication, we can assess their effectiveness as a guiding tool for designing human-centered co-creative AI communication and identify opportunities for improvement. Future research can also explore how individual preferences through the configuration tool might influence the user experience to identify more nuanced human-centered design implications.

There is also an opportunity for FAICO to connect more explicitly with the unique challenges of co-creative processes, such as whether different types of communication can spark opportunities for reflection \cite{ford_reflection_across} or moments of surprise \cite{xaixarts_workshop}. In particular, there is a need to explore the trade-offs in AI communication design identified in FAICO and related work — such as how to balance additional multimodal communication elements alongside the risk of overwhelming users and breaking their creative flow \cite{csikszentmihalyi_flow_1990}.

Additionally, we want to identify trends and gaps in the AI communication landscape by analyzing a large corpus of existing co-creative AI systems across various domains. This analysis will provide valuable insights into current practices, highlight best approaches, and uncover gaps in the design of AI communication. The results will guide future research directions and help establish a roadmap for advancing the field of human-AI co-creativity. Overall, this paper lays preliminary groundwork to guide the design of AI communication and to help improving future co-creative AI systems.

\bibliographystyle{ACM-Reference-Format}
\bibliography{sample-base}


\begin{thebibliography}{107}


\ifx \showCODEN    \undefined \def \showCODEN     #1{\unskip}     \fi
\ifx \showDOI      \undefined \def \showDOI       #1{#1}\fi
\ifx \showISBNx    \undefined \def \showISBNx     #1{\unskip}     \fi
\ifx \showISBNxiii \undefined \def \showISBNxiii  #1{\unskip}     \fi
\ifx \showISSN     \undefined \def \showISSN      #1{\unskip}     \fi
\ifx \showLCCN     \undefined \def \showLCCN      #1{\unskip}     \fi
\ifx \shownote     \undefined \def \shownote      #1{#1}          \fi
\ifx \showarticletitle \undefined \def \showarticletitle #1{#1}   \fi
\ifx \showURL      \undefined \def \showURL       {\relax}        \fi
\providecommand\bibfield[2]{#2}
\providecommand\bibinfo[2]{#2}
\providecommand\natexlab[1]{#1}
\providecommand\showeprint[2][]{arXiv:#2}

\bibitem[ope(linea)]%
        {openaiChatGPTOptimizing}
 \bibinfo{year}{Online}\natexlab{a}.
\newblock \bibinfo{booktitle}{\emph{{C}hat{G}{P}{T}: {O}ptimizing {L}anguage {M}odels for {D}ialogue --- openai.com}}.
\newblock
\urldef\tempurl%
\url{https://openai.com/blog/chatgpt/}
\showURL{%
\tempurl}


\bibitem[ope(lineb)]%
        {openaiDALLE}
 \bibinfo{year}{Online}\natexlab{b}.
\newblock \bibinfo{booktitle}{\emph{{D}{A}{L}{L}·{E} 2 --- openai.com}}.
\newblock
\urldef\tempurl%
\url{https://openai.com/dall-e-2/}
\showURL{%
\tempurl}


\bibitem[mid(line)]%
        {midjourney}
 \bibinfo{year}{Online}\natexlab{}.
\newblock \bibinfo{booktitle}{\emph{{M}idjourney}}.
\newblock
\urldef\tempurl%
\url{https://www.midjourney.com/}
\showURL{%
\tempurl}


\bibitem[Anzabi and Umemuro(2023)]%
        {anzabi2023effect}
\bibfield{author}{\bibinfo{person}{Naeimeh Anzabi} {and} \bibinfo{person}{Hiroyuki Umemuro}.} \bibinfo{year}{2023}\natexlab{}.
\newblock \showarticletitle{The Effect of Social Robots' Listening Behaviors on Animacy, Likeability and Perceived Intelligence}. In \bibinfo{booktitle}{\emph{Proceedings of the 11th International Conference on Human-Agent Interaction}}. \bibinfo{pages}{343--350}.
\newblock


\bibitem[Bente et~al\mbox{.}(2004)]%
        {bente2004social}
\bibfield{author}{\bibinfo{person}{Gary Bente}, \bibinfo{person}{Sabine R{\"u}ggenberg}, {and} \bibinfo{person}{Nicole~C Kr{\"a}mer}.} \bibinfo{year}{2004}\natexlab{}.
\newblock \showarticletitle{Social presence and interpersonal trust in avatar-based, collaborative net-communications}. In \bibinfo{booktitle}{\emph{Proceedings of the Seventh Annual International Workshop on Presence}}. \bibinfo{pages}{54--61}.
\newblock


\bibitem[Berry(2006)]%
        {berry2006can}
\bibfield{author}{\bibinfo{person}{Gregory~R Berry}.} \bibinfo{year}{2006}\natexlab{}.
\newblock \showarticletitle{Can computer-mediated asynchronous communication improve team processes and decision making? Learning from the management literature}.
\newblock \bibinfo{journal}{\emph{The Journal of Business Communication (1973)}} \bibinfo{volume}{43}, \bibinfo{number}{4} (\bibinfo{year}{2006}), \bibinfo{pages}{344--366}.
\newblock


\bibitem[Berry(2024)]%
        {berry2024reactive}
\bibfield{author}{\bibinfo{person}{Jacquelyn Berry}.} \bibinfo{year}{2024}\natexlab{}.
\newblock \showarticletitle{Reactive Ai Feedback Improves Task Performance Over Time}.
\newblock \bibinfo{journal}{\emph{Available at SSRN 4788587}} (\bibinfo{year}{2024}).
\newblock


\bibitem[Bown(2015)]%
        {bown2015player}
\bibfield{author}{\bibinfo{person}{Oliver Bown}.} \bibinfo{year}{2015}\natexlab{}.
\newblock \showarticletitle{Player Responses to a Live Algorithm: Conceptualising computational creativity without recourse to human comparisons?}. In \bibinfo{booktitle}{\emph{ICCC}}. \bibinfo{pages}{126--133}.
\newblock


\bibitem[Bown et~al\mbox{.}(2020)]%
        {bown2020speculative}
\bibfield{author}{\bibinfo{person}{Oliver Bown}, \bibinfo{person}{Kazjon Grace}, \bibinfo{person}{Liam Bray}, {and} \bibinfo{person}{Dan Ventura}.} \bibinfo{year}{2020}\natexlab{}.
\newblock \showarticletitle{Speculative Exploration of the Role of Dialogue in Human-ComputerCo-creation.}. In \bibinfo{booktitle}{\emph{ICCC}}. \bibinfo{pages}{25--32}.
\newblock


\bibitem[Bryan-Kinns et~al\mbox{.}(2023)]%
        {xaixarts_workshop}
\bibfield{author}{\bibinfo{person}{Nick Bryan-Kinns}, \bibinfo{person}{Corey Ford}, \bibinfo{person}{Alan Chamberlain}, \bibinfo{person}{Steven~David Benford}, \bibinfo{person}{Helen Kennedy}, \bibinfo{person}{Zijin Li}, \bibinfo{person}{Wu Qiong}, \bibinfo{person}{Gus~G. Xia}, {and} \bibinfo{person}{Jeba Rezwana}.} \bibinfo{year}{2023}\natexlab{}.
\newblock \showarticletitle{Explainable AI for the Arts: XAIxArts}. In \bibinfo{booktitle}{\emph{Proceedings of the 15th Conference on Creativity and Cognition}} (Virtual Event, USA) \emph{(\bibinfo{series}{C\&C '23})}. \bibinfo{publisher}{Association for Computing Machinery}, \bibinfo{address}{New York, NY, USA}, \bibinfo{pages}{1–7}.
\newblock
\showISBNx{9798400701801}
\urldef\tempurl%
\url{https://doi.org/10.1145/3591196.3593517}
\showDOI{\tempurl}


\bibitem[Bryan-Kinns et~al\mbox{.}(2024a)]%
        {xaixarts_workshop_2}
\bibfield{author}{\bibinfo{person}{Nick Bryan-Kinns}, \bibinfo{person}{Corey Ford}, \bibinfo{person}{Shuoyang Zheng}, \bibinfo{person}{Helen Kennedy}, \bibinfo{person}{Alan Chamberlain}, \bibinfo{person}{Makayla Lewis}, \bibinfo{person}{Drew Hemment}, \bibinfo{person}{Zijin Li}, \bibinfo{person}{Qiong Wu}, \bibinfo{person}{Lanxi Xiao}, \bibinfo{person}{Gus Xia}, \bibinfo{person}{Jeba Rezwana}, \bibinfo{person}{Michael Clemens}, {and} \bibinfo{person}{Gabriel Vigliensoni}.} \bibinfo{year}{2024}\natexlab{a}.
\newblock \showarticletitle{Explainable AI for the Arts 2 (XAIxArts2)}. In \bibinfo{booktitle}{\emph{Proceedings of the 16th Conference on Creativity \& Cognition}} (Chicago, IL, USA) \emph{(\bibinfo{series}{C\&C '24})}. \bibinfo{publisher}{Association for Computing Machinery}, \bibinfo{address}{New York, NY, USA}, \bibinfo{pages}{86–92}.
\newblock
\showISBNx{9798400704857}
\urldef\tempurl%
\url{https://doi.org/10.1145/3635636.3660763}
\showDOI{\tempurl}


\bibitem[Bryan-Kinns and Hamilton(2012)]%
        {bryan2012identifying}
\bibfield{author}{\bibinfo{person}{Nick Bryan-Kinns} {and} \bibinfo{person}{Fraser Hamilton}.} \bibinfo{year}{2012}\natexlab{}.
\newblock \showarticletitle{Identifying mutual engagement}.
\newblock \bibinfo{journal}{\emph{Behaviour \& Information Technology}} \bibinfo{volume}{31}, \bibinfo{number}{2} (\bibinfo{year}{2012}), \bibinfo{pages}{101--125}.
\newblock


\bibitem[Bryan-Kinns et~al\mbox{.}(2024b)]%
        {using_incongrous}
\bibfield{author}{\bibinfo{person}{Nick Bryan-Kinns}, \bibinfo{person}{Ashley Noel-Hirst}, {and} \bibinfo{person}{Corey Ford}.} \bibinfo{year}{2024}\natexlab{b}.
\newblock \showarticletitle{Using Incongruous Genres to Explore Music Making with AI Generated Content}. In \bibinfo{booktitle}{\emph{Proceedings of the 16th Conference on Creativity \& Cognition}} (Chicago, IL, USA) \emph{(\bibinfo{series}{C\&C '24})}. \bibinfo{publisher}{Association for Computing Machinery}, \bibinfo{address}{New York, NY, USA}, \bibinfo{pages}{229–240}.
\newblock
\showISBNx{9798400704857}
\urldef\tempurl%
\url{https://doi.org/10.1145/3635636.3656198}
\showDOI{\tempurl}


\bibitem[Colley et~al\mbox{.}(2021)]%
        {colley2021}
\bibfield{author}{\bibinfo{person}{Mark Colley}, \bibinfo{person}{Jan~Henry Belz}, {and} \bibinfo{person}{Enrico Rukzio}.} \bibinfo{year}{2021}\natexlab{}.
\newblock \showarticletitle{Investigating the Effects of Feedback Communication of Autonomous Vehicles}. In \bibinfo{booktitle}{\emph{13th International Conference on Automotive User Interfaces and Interactive Vehicular Applications}} (Leeds, United Kingdom) \emph{(\bibinfo{series}{AutomotiveUI '21})}. \bibinfo{publisher}{Association for Computing Machinery}, \bibinfo{address}{New York, NY, USA}, \bibinfo{pages}{263–273}.
\newblock
\showISBNx{9781450380638}
\urldef\tempurl%
\url{https://doi.org/10.1145/3409118.3475133}
\showDOI{\tempurl}


\bibitem[Csíkszentmihályi(1990)]%
        {csikszentmihalyi_flow_1990}
\bibfield{author}{\bibinfo{person}{Mihály Csíkszentmihályi}.} \bibinfo{year}{1990}\natexlab{}.
\newblock \bibinfo{booktitle}{\emph{Flow: {The} {Psychology} of {Optimal} {Experience}}}.
\newblock \bibinfo{publisher}{Harper Collins}, \bibinfo{address}{New York, USA}.
\newblock


\bibitem[Dafoe et~al\mbox{.}(2021)]%
        {dafoe2021cooperative}
\bibfield{author}{\bibinfo{person}{Allan Dafoe}, \bibinfo{person}{Yoram Bachrach}, \bibinfo{person}{Gillian Hadfield}, \bibinfo{person}{Eric Horvitz}, \bibinfo{person}{Kate Larson}, {and} \bibinfo{person}{Thore Graepel}.} \bibinfo{year}{2021}\natexlab{}.
\newblock \showarticletitle{Cooperative AI: machines must learn to find common ground}.
\newblock \bibinfo{journal}{\emph{Nature}} \bibinfo{volume}{593}, \bibinfo{number}{7857} (\bibinfo{year}{2021}), \bibinfo{pages}{33--36}.
\newblock


\bibitem[Davis(2013)]%
        {davis2013human}
\bibfield{author}{\bibinfo{person}{Nicholas Davis}.} \bibinfo{year}{2013}\natexlab{}.
\newblock \showarticletitle{Human-computer co-creativity: Blending human and computational creativity}. In \bibinfo{booktitle}{\emph{Proceedings of the AAAI Conference on Artificial Intelligence and Interactive Digital Entertainment}}, Vol.~\bibinfo{volume}{9}. \bibinfo{pages}{9--12}.
\newblock


\bibitem[Davis et~al\mbox{.}(2016)]%
        {davis2016empirically}
\bibfield{author}{\bibinfo{person}{Nicholas Davis}, \bibinfo{person}{Chih-PIn Hsiao}, \bibinfo{person}{Kunwar Yashraj~Singh}, \bibinfo{person}{Lisa Li}, {and} \bibinfo{person}{Brian Magerko}.} \bibinfo{year}{2016}\natexlab{}.
\newblock \showarticletitle{Empirically studying participatory sense-making in abstract drawing with a co-creative cognitive agent}. In \bibinfo{booktitle}{\emph{Proceedings of the 21st International Conference on Intelligent User Interfaces}}. \bibinfo{pages}{196--207}.
\newblock


\bibitem[Dev and Camp(2020)]%
        {dev2020user}
\bibfield{author}{\bibinfo{person}{Jayati Dev} {and} \bibinfo{person}{L~Jean Camp}.} \bibinfo{year}{2020}\natexlab{}.
\newblock \showarticletitle{User engagement with chatbots: a discursive psychology approach}. In \bibinfo{booktitle}{\emph{Proceedings of the 2nd Conference on Conversational User Interfaces}}. \bibinfo{pages}{1--4}.
\newblock


\bibitem[D'mello and Graesser(2013)]%
        {d2013autotutor}
\bibfield{author}{\bibinfo{person}{Sidney D'mello} {and} \bibinfo{person}{Art Graesser}.} \bibinfo{year}{2013}\natexlab{}.
\newblock \showarticletitle{AutoTutor and affective AutoTutor: Learning by talking with cognitively and emotionally intelligent computers that talk back}.
\newblock \bibinfo{journal}{\emph{ACM Transactions on Interactive Intelligent Systems (TiiS)}} \bibinfo{volume}{2}, \bibinfo{number}{4} (\bibinfo{year}{2013}), \bibinfo{pages}{1--39}.
\newblock


\bibitem[Ehsan et~al\mbox{.}(2021)]%
        {ehsan2021expanding}
\bibfield{author}{\bibinfo{person}{Upol Ehsan}, \bibinfo{person}{Q~Vera Liao}, \bibinfo{person}{Michael Muller}, \bibinfo{person}{Mark~O Riedl}, {and} \bibinfo{person}{Justin~D Weisz}.} \bibinfo{year}{2021}\natexlab{}.
\newblock \showarticletitle{Expanding explainability: Towards social transparency in ai systems}. In \bibinfo{booktitle}{\emph{Proceedings of the 2021 CHI Conference on Human Factors in Computing Systems}}. \bibinfo{pages}{1--19}.
\newblock


\bibitem[El-Zanfaly et~al\mbox{.}(2022)]%
        {el2022sand}
\bibfield{author}{\bibinfo{person}{Dina El-Zanfaly}, \bibinfo{person}{Yiwei Huang}, {and} \bibinfo{person}{Yanwen Dong}.} \bibinfo{year}{2022}\natexlab{}.
\newblock \showarticletitle{Sand Playground: Designing Human-AI physical Interface for Co-creation in Motion}. In \bibinfo{booktitle}{\emph{Proceedings of the 14th Conference on Creativity and Cognition}}. \bibinfo{pages}{49--55}.
\newblock


\bibitem[Endrass et~al\mbox{.}(2009)]%
        {endrass2009culture}
\bibfield{author}{\bibinfo{person}{Birgit Endrass}, \bibinfo{person}{Matthias Rehm}, {and} \bibinfo{person}{Elisabeth Andr\'{e}}.} \bibinfo{year}{2009}\natexlab{}.
\newblock \showarticletitle{Culture-specific communication management for virtual agents}. In \bibinfo{booktitle}{\emph{Proceedings of The 8th International Conference on Autonomous Agents and Multiagent Systems - Volume 1}} (Budapest, Hungary) \emph{(\bibinfo{series}{AAMAS '09})}. \bibinfo{publisher}{International Foundation for Autonomous Agents and Multiagent Systems}, \bibinfo{address}{Richland, SC}, \bibinfo{pages}{281–287}.
\newblock
\showISBNx{9780981738161}


\bibitem[Esau-Held et~al\mbox{.}(2023)]%
        {esau2023foggy}
\bibfield{author}{\bibinfo{person}{Margarita Esau-Held}, \bibinfo{person}{Andrew Marsh}, \bibinfo{person}{Veronika Krau{\ss}}, {and} \bibinfo{person}{Gunnar Stevens}.} \bibinfo{year}{2023}\natexlab{}.
\newblock \showarticletitle{“Foggy sounds like nothing”—enriching the experience of voice assistants with sonic overlays}.
\newblock \bibinfo{journal}{\emph{Personal and Ubiquitous Computing}} \bibinfo{volume}{27}, \bibinfo{number}{5} (\bibinfo{year}{2023}), \bibinfo{pages}{1927--1947}.
\newblock


\bibitem[Fan et~al\mbox{.}(2022)]%
        {fan2022human}
\bibfield{author}{\bibinfo{person}{Mingming Fan}, \bibinfo{person}{Xianyou Yang}, \bibinfo{person}{TszTung Yu}, \bibinfo{person}{Q~Vera Liao}, {and} \bibinfo{person}{Jian Zhao}.} \bibinfo{year}{2022}\natexlab{}.
\newblock \bibinfo{title}{Human-AI Collaboration for UX Evaluation: Effects of Explanation and Synchronization. 6 (2022), 96: 1--96: 32. Issue CSCW1}.
\newblock
\newblock


\bibitem[Ford et~al\mbox{.}(2024)]%
        {ford_reflection_across}
\bibfield{author}{\bibinfo{person}{Corey Ford}, \bibinfo{person}{Ashley Noel-Hirst}, \bibinfo{person}{Sara Cardinale}, \bibinfo{person}{Jackson Loth}, \bibinfo{person}{Pedro Sarmento}, \bibinfo{person}{Elizabeth Wilson}, \bibinfo{person}{Lewis Wolstanholme}, \bibinfo{person}{Kyle Worrall}, {and} \bibinfo{person}{Nick Bryan-Kinns}.} \bibinfo{year}{2024}\natexlab{}.
\newblock \showarticletitle{Reflection Across AI-based Music Composition}. In \bibinfo{booktitle}{\emph{Proceedings of the 16th Conference on Creativity \& Cognition}} (Chicago, IL, USA) \emph{(\bibinfo{series}{C\&C '24})}. \bibinfo{publisher}{Association for Computing Machinery}, \bibinfo{address}{New York, NY, USA}, \bibinfo{pages}{398–412}.
\newblock
\showISBNx{9798400704857}
\urldef\tempurl%
\url{https://doi.org/10.1145/3635636.3656185}
\showDOI{\tempurl}


\bibitem[Franke and Piller(2004)]%
        {franke2004value}
\bibfield{author}{\bibinfo{person}{Nikolaus Franke} {and} \bibinfo{person}{Frank Piller}.} \bibinfo{year}{2004}\natexlab{}.
\newblock \showarticletitle{Value creation by toolkits for user innovation and design: The case of the watch market}.
\newblock \bibinfo{journal}{\emph{Journal of product innovation management}} \bibinfo{volume}{21}, \bibinfo{number}{6} (\bibinfo{year}{2004}), \bibinfo{pages}{401--415}.
\newblock


\bibitem[French(2000)]%
        {french2000turing}
\bibfield{author}{\bibinfo{person}{Robert~M French}.} \bibinfo{year}{2000}\natexlab{}.
\newblock \showarticletitle{The Turing Test: the first 50 years}.
\newblock \bibinfo{journal}{\emph{Trends in cognitive sciences}} \bibinfo{volume}{4}, \bibinfo{number}{3} (\bibinfo{year}{2000}), \bibinfo{pages}{115--122}.
\newblock


\bibitem[Frich et~al\mbox{.}(2019)]%
        {frich2019}
\bibfield{author}{\bibinfo{person}{Jonas Frich}, \bibinfo{person}{Lindsay MacDonald~Vermeulen}, \bibinfo{person}{Christian Remy}, \bibinfo{person}{Michael~Mose Biskjaer}, {and} \bibinfo{person}{Peter Dalsgaard}.} \bibinfo{year}{2019}\natexlab{}.
\newblock \showarticletitle{Mapping the Landscape of Creativity Support Tools in HCI}. In \bibinfo{booktitle}{\emph{Proceedings of the 2019 CHI Conference on Human Factors in Computing Systems}} (Glasgow, Scotland Uk) \emph{(\bibinfo{series}{CHI '19})}. \bibinfo{publisher}{Association for Computing Machinery}, \bibinfo{address}{New York, NY, USA}, \bibinfo{pages}{1–18}.
\newblock
\showISBNx{9781450359702}
\urldef\tempurl%
\url{https://doi.org/10.1145/3290605.3300619}
\showDOI{\tempurl}


\bibitem[Garibay et~al\mbox{.}(2023)]%
        {ozmen2023sixchallenges}
\bibfield{author}{\bibinfo{person}{Ozlem~Ozmen Garibay}, \bibinfo{person}{Brent Winslow}, \bibinfo{person}{Salvatore Andolina}, \bibinfo{person}{Margherita Antona}, \bibinfo{person}{Anja Bodenschatz}, \bibinfo{person}{Constantinos Coursaris}, \bibinfo{person}{Gregory Falco}, \bibinfo{person}{Stephen~M. Fiore}, \bibinfo{person}{Ivan Garibay}, \bibinfo{person}{Keri Grieman}, \bibinfo{person}{John~C. Havens}, \bibinfo{person}{Marina Jirotka}, \bibinfo{person}{Hernisa Kacorri}, \bibinfo{person}{Waldemar Karwowski}, \bibinfo{person}{Joe Kider}, \bibinfo{person}{Joseph Konstan}, \bibinfo{person}{Sean Koon}, \bibinfo{person}{Monica Lopez-Gonzalez}, \bibinfo{person}{Iliana Maifeld-Carucci}, \bibinfo{person}{Sean McGregor}, \bibinfo{person}{Gavriel Salvendy}, \bibinfo{person}{Ben Shneiderman}, \bibinfo{person}{Constantine Stephanidis}, \bibinfo{person}{Christina Strobel}, \bibinfo{person}{Carolyn~Ten Holter}, {and} \bibinfo{person}{Wei Xu}.} \bibinfo{year}{2023}\natexlab{}.
\newblock \showarticletitle{Six Human-Centered Artificial Intelligence Grand Challenges}.
\newblock \bibinfo{journal}{\emph{International Journal of Human–Computer Interaction}} \bibinfo{volume}{39}, \bibinfo{number}{3} (\bibinfo{year}{2023}), \bibinfo{pages}{391--437}.
\newblock
\urldef\tempurl%
\url{https://doi.org/10.1080/10447318.2022.2153320}
\showDOI{\tempurl}


\bibitem[Gilad et~al\mbox{.}(2021)]%
        {gilad2021warmth}
\bibfield{author}{\bibinfo{person}{Zohar Gilad}, \bibinfo{person}{Ofra Amir}, {and} \bibinfo{person}{Liat Levontin}.} \bibinfo{year}{2021}\natexlab{}.
\newblock \showarticletitle{The Effects of Warmth and Competence Perceptions on Users' Choice of an AI System}. In \bibinfo{booktitle}{\emph{Proceedings of the 2021 CHI Conference on Human Factors in Computing Systems}} (Yokohama, Japan) \emph{(\bibinfo{series}{CHI '21})}. \bibinfo{publisher}{Association for Computing Machinery}, \bibinfo{address}{New York, NY, USA}, Article \bibinfo{articleno}{583}, \bibinfo{numpages}{13}~pages.
\newblock
\showISBNx{9781450380966}
\urldef\tempurl%
\url{https://doi.org/10.1145/3411764.3446863}
\showDOI{\tempurl}


\bibitem[Gunning(2016)]%
        {gunning_2016}
\bibfield{author}{\bibinfo{person}{David Gunning}.} \bibinfo{year}{2016}\natexlab{}.
\newblock \showarticletitle{Explainable {A}rtificial {I}ntelligence ({XAI})}.
\newblock \bibinfo{journal}{\emph{DARPA/I2O Proposers Day}} (\bibinfo{date}{Aug} \bibinfo{year}{2016}).
\newblock


\bibitem[Guzdial and Riedl(2019)]%
        {Guzdial2019}
\bibfield{author}{\bibinfo{person}{Matthew Guzdial} {and} \bibinfo{person}{Mark Riedl}.} \bibinfo{year}{2019}\natexlab{}.
\newblock \showarticletitle{An interaction framework for studying co-creative ai}.
\newblock \bibinfo{journal}{\emph{arXiv preprint arXiv:1903.09709}} (\bibinfo{year}{2019}).
\newblock


\bibitem[Guzman and Lewis(2020)]%
        {guzman2020artificial}
\bibfield{author}{\bibinfo{person}{Andrea~L Guzman} {and} \bibinfo{person}{Seth~C Lewis}.} \bibinfo{year}{2020}\natexlab{}.
\newblock \showarticletitle{Artificial intelligence and communication: A human--machine communication research agenda}.
\newblock \bibinfo{journal}{\emph{New media \& society}} \bibinfo{volume}{22}, \bibinfo{number}{1} (\bibinfo{year}{2020}), \bibinfo{pages}{70--86}.
\newblock


\bibitem[Hancock et~al\mbox{.}(2020)]%
        {hancock2020ai}
\bibfield{author}{\bibinfo{person}{Jeffrey~T Hancock}, \bibinfo{person}{Mor Naaman}, {and} \bibinfo{person}{Karen Levy}.} \bibinfo{year}{2020}\natexlab{}.
\newblock \showarticletitle{AI-mediated communication: Definition, research agenda, and ethical considerations}.
\newblock \bibinfo{journal}{\emph{Journal of Computer-Mediated Communication}} \bibinfo{volume}{25}, \bibinfo{number}{1} (\bibinfo{year}{2020}), \bibinfo{pages}{89--100}.
\newblock


\bibitem[Harboe and Huang(2015)]%
        {harboe2015affinity}
\bibfield{author}{\bibinfo{person}{Gunnar Harboe} {and} \bibinfo{person}{Elaine~M. Huang}.} \bibinfo{year}{2015}\natexlab{}.
\newblock \showarticletitle{Real-World Affinity Diagramming Practices: Bridging the Paper-Digital Gap}. In \bibinfo{booktitle}{\emph{Proceedings of the 33rd Annual ACM Conference on Human Factors in Computing Systems}} (Seoul, Republic of Korea) \emph{(\bibinfo{series}{CHI '15})}. \bibinfo{publisher}{Association for Computing Machinery}, \bibinfo{address}{New York, NY, USA}, \bibinfo{pages}{95–104}.
\newblock
\showISBNx{9781450331456}
\urldef\tempurl%
\url{https://doi.org/10.1145/2702123.2702561}
\showDOI{\tempurl}


\bibitem[Healey(2021)]%
        {healey2021human}
\bibfield{author}{\bibinfo{person}{Patrick~GT Healey}.} \bibinfo{year}{2021}\natexlab{}.
\newblock \showarticletitle{Human-Like Communication}.
\newblock \bibinfo{journal}{\emph{Oxford University Press, Oxford, England}} (\bibinfo{year}{2021}).
\newblock


\bibitem[Healey et~al\mbox{.}(2005)]%
        {healey2005inter}
\bibfield{author}{\bibinfo{person}{Patrick~GT Healey}, \bibinfo{person}{Joe Leach}, {and} \bibinfo{person}{Nick Bryan-Kinns}.} \bibinfo{year}{2005}\natexlab{}.
\newblock \showarticletitle{Inter-play: Understanding group music improvisation as a form of everyday interaction}.
\newblock \bibinfo{journal}{\emph{Proceedings of Less is More—Simple Computing in an Age of Complexity}} (\bibinfo{year}{2005}).
\newblock


\bibitem[Hsieh et~al\mbox{.}(2023)]%
        {hsieh2023cards}
\bibfield{author}{\bibinfo{person}{Gary Hsieh}, \bibinfo{person}{Brett~A Halperin}, \bibinfo{person}{Evan Schmitz}, \bibinfo{person}{Yen~Nee Chew}, {and} \bibinfo{person}{Yuan-Chi Tseng}.} \bibinfo{year}{2023}\natexlab{}.
\newblock \showarticletitle{What is in the cards: Exploring uses, patterns, and trends in design cards}. In \bibinfo{booktitle}{\emph{Proceedings of the 2023 CHI Conference on Human Factors in Computing Systems}}. \bibinfo{pages}{1--18}.
\newblock


\bibitem[Huang et~al\mbox{.}(2021)]%
        {huang2021women}
\bibfield{author}{\bibinfo{person}{Yan Huang}, \bibinfo{person}{S~Shyam Sundar}, \bibinfo{person}{Zhiyao Ye}, {and} \bibinfo{person}{Ariel~Celeste Johnson}.} \bibinfo{year}{2021}\natexlab{}.
\newblock \showarticletitle{Do women and extroverts perceive interactivity differently than men and introverts? Role of individual differences in responses to HCI vs. CMC interactivity}.
\newblock \bibinfo{journal}{\emph{Computers in Human Behavior}}  \bibinfo{volume}{123} (\bibinfo{year}{2021}), \bibinfo{pages}{106881}.
\newblock


\bibitem[Hui and Boutilier(2008)]%
        {hui2008toward}
\bibfield{author}{\bibinfo{person}{Bowen Hui} {and} \bibinfo{person}{Craig Boutilier}.} \bibinfo{year}{2008}\natexlab{}.
\newblock \showarticletitle{Toward experiential utility elicitation for interface customization}. In \bibinfo{booktitle}{\emph{Proceedings of the Twenty-Fourth Conference on Uncertainty in Artificial Intelligence}}. \bibinfo{pages}{298--305}.
\newblock


\bibitem[Hwang(2022)]%
        {hwang2022too}
\bibfield{author}{\bibinfo{person}{Angel Hsing-Chi Hwang}.} \bibinfo{year}{2022}\natexlab{}.
\newblock \showarticletitle{Too late to be creative? AI-empowered tools in creative processes}. In \bibinfo{booktitle}{\emph{CHI conference on human factors in computing systems extended abstracts}}. \bibinfo{pages}{1--9}.
\newblock


\bibitem[Jain et~al\mbox{.}(2023)]%
        {jain2023co}
\bibfield{author}{\bibinfo{person}{Pranut Jain}, \bibinfo{person}{Rosta Farzan}, {and} \bibinfo{person}{Adam~J Lee}.} \bibinfo{year}{2023}\natexlab{}.
\newblock \showarticletitle{Co-Designing with Users the Explanations for a Proactive Auto-Response Messaging Agent}.
\newblock \bibinfo{journal}{\emph{Proceedings of the ACM on Human-Computer Interaction}} \bibinfo{volume}{7}, \bibinfo{number}{MHCI} (\bibinfo{year}{2023}), \bibinfo{pages}{1--23}.
\newblock


\bibitem[Janaka et~al\mbox{.}(2023)]%
        {nanayakkarawasam_icon_2023}
\bibfield{author}{\bibinfo{person}{Nuwan Nanayakkarawasam Peru~Kandage Janaka}, \bibinfo{person}{Shengdong Zhao}, {and} \bibinfo{person}{Shardul Sapkota}.} \bibinfo{year}{2023}\natexlab{}.
\newblock \showarticletitle{Can Icons Outperform Text? Understanding the Role of Pictograms in OHMD Notifications}. In \bibinfo{booktitle}{\emph{Proceedings of the 2023 CHI Conference on Human Factors in Computing Systems}} (Hamburg, Germany) \emph{(\bibinfo{series}{CHI '23})}. \bibinfo{publisher}{Association for Computing Machinery}, \bibinfo{address}{New York, NY, USA}, Article \bibinfo{articleno}{575}, \bibinfo{numpages}{23}~pages.
\newblock
\showISBNx{9781450394215}
\urldef\tempurl%
\url{https://doi.org/10.1145/3544548.3580891}
\showDOI{\tempurl}


\bibitem[Kantosalo et~al\mbox{.}(2020)]%
        {kantosalo2020modalities}
\bibfield{author}{\bibinfo{person}{Anna Kantosalo}, \bibinfo{person}{Prashanth~Thattai Ravikumar}, \bibinfo{person}{Kazjon Grace}, {and} \bibinfo{person}{Tapio Takala}.} \bibinfo{year}{2020}\natexlab{}.
\newblock \showarticletitle{Modalities, Styles and Strategies: An Interaction Framework for Human-Computer Co-Creativity.}. In \bibinfo{booktitle}{\emph{International Conference on Computational Creativity}}. \bibinfo{pages}{57--64}.
\newblock


\bibitem[Kantosalo et~al\mbox{.}(2014)]%
        {kantosalo2014isolation}
\bibfield{author}{\bibinfo{person}{Anna Kantosalo}, \bibinfo{person}{Jukka~M Toivanen}, \bibinfo{person}{Ping Xiao}, {and} \bibinfo{person}{Hannu Toivonen}.} \bibinfo{year}{2014}\natexlab{}.
\newblock \showarticletitle{From Isolation to Involvement: Adapting Machine Creativity Software to Support Human-Computer Co-Creation.}. In \bibinfo{booktitle}{\emph{ICCC}}. \bibinfo{pages}{1--7}.
\newblock


\bibitem[Karimi et~al\mbox{.}(2020)]%
        {karimi2020creative}
\bibfield{author}{\bibinfo{person}{Pegah Karimi}, \bibinfo{person}{Jeba Rezwana}, \bibinfo{person}{Safat Siddiqui}, \bibinfo{person}{Mary~Lou Maher}, {and} \bibinfo{person}{Nasrin Dehbozorgi}.} \bibinfo{year}{2020}\natexlab{}.
\newblock \showarticletitle{Creative sketching partner: an analysis of human-AI co-creativity}. In \bibinfo{booktitle}{\emph{Proceedings of the 25th International Conference on Intelligent User Interfaces}}. \bibinfo{pages}{221--230}.
\newblock


\bibitem[Karray et~al\mbox{.}(2008)]%
        {karray2008human}
\bibfield{author}{\bibinfo{person}{Fakhreddine Karray}, \bibinfo{person}{Milad Alemzadeh}, \bibinfo{person}{Jamil Abou~Saleh}, {and} \bibinfo{person}{Mo~Nours Arab}.} \bibinfo{year}{2008}\natexlab{}.
\newblock \showarticletitle{Human-computer interaction: Overview on state of the art}.
\newblock \bibinfo{journal}{\emph{International journal on smart sensing and intelligent systems}} \bibinfo{volume}{1}, \bibinfo{number}{1} (\bibinfo{year}{2008}), \bibinfo{pages}{137--159}.
\newblock


\bibitem[Kim et~al\mbox{.}(2021)]%
        {kim2021ai}
\bibfield{author}{\bibinfo{person}{Jihyun Kim}, \bibinfo{person}{Kelly Merrill~Jr}, {and} \bibinfo{person}{Chad Collins}.} \bibinfo{year}{2021}\natexlab{}.
\newblock \showarticletitle{AI as a friend or assistant: The mediating role of perceived usefulness in social AI vs. functional AI}.
\newblock \bibinfo{journal}{\emph{Telematics and Informatics}}  \bibinfo{volume}{64} (\bibinfo{year}{2021}), \bibinfo{pages}{101694}.
\newblock


\bibitem[Kozierok et~al\mbox{.}(2021)]%
        {DARPA}
\bibfield{author}{\bibinfo{person}{Robyn Kozierok}, \bibinfo{person}{John Aberdeen}, \bibinfo{person}{Cheryl Clark}, \bibinfo{person}{Christopher Garay}, \bibinfo{person}{Bradley Goodman}, \bibinfo{person}{Tonia Korves}, \bibinfo{person}{Lynette Hirschman}, \bibinfo{person}{Patricia~L. McDermott}, {and} \bibinfo{person}{Matthew~W. Peterson}.} \bibinfo{year}{2021}\natexlab{}.
\newblock \showarticletitle{Hallmarks of {H}uman-{M}achine {C}ollaboration: A framework for assessment in the {DARPA} Communicating with {C}omputers {P}rogram}.
\newblock  (\bibinfo{year}{2021}).
\newblock
\showeprint[arxiv]{2102.04958}~[cs.HC]


\bibitem[Kraus et~al\mbox{.}(2022)]%
        {kraus2022kurt}
\bibfield{author}{\bibinfo{person}{Matthias Kraus}, \bibinfo{person}{Nicolas Wagner}, \bibinfo{person}{Wolfgang Minker}, \bibinfo{person}{Ankita Agrawal}, \bibinfo{person}{Artur Schmidt}, \bibinfo{person}{Pranav~Krishna Prasad}, {and} \bibinfo{person}{Wolfgang Ertel}.} \bibinfo{year}{2022}\natexlab{}.
\newblock \showarticletitle{KURT: A household assistance robot capable of proactive dialogue}. In \bibinfo{booktitle}{\emph{2022 17th ACM/IEEE International Conference on Human-Robot Interaction (HRI)}}. IEEE, \bibinfo{pages}{855--859}.
\newblock


\bibitem[Leong and Sung(2023)]%
        {leong2023exploratory}
\bibfield{author}{\bibinfo{person}{Kelvin Leong} {and} \bibinfo{person}{Anna Sung}.} \bibinfo{year}{2023}\natexlab{}.
\newblock \showarticletitle{An Exploratory Study of How Emotion Tone Presented in A Message Influences Artificial Intelligence (AI) Powered Recommendation System}.
\newblock  (\bibinfo{year}{2023}).
\newblock


\bibitem[Liang et~al\mbox{.}(2019)]%
        {liang2019implicit}
\bibfield{author}{\bibinfo{person}{Claire Liang}, \bibinfo{person}{Julia Proft}, \bibinfo{person}{Erik Andersen}, {and} \bibinfo{person}{Ross~A Knepper}.} \bibinfo{year}{2019}\natexlab{}.
\newblock \showarticletitle{Implicit communication of actionable information in human-ai teams}. In \bibinfo{booktitle}{\emph{Proceedings of the 2019 CHI Conference on Human Factors in Computing Systems}}. \bibinfo{pages}{1--13}.
\newblock


\bibitem[Liapis et~al\mbox{.}(2014)]%
        {liapis2014computational}
\bibfield{author}{\bibinfo{person}{Antonios Liapis}, \bibinfo{person}{Georgios~N Yannakakis}, {and} \bibinfo{person}{Julian Togelius}.} \bibinfo{year}{2014}\natexlab{}.
\newblock \showarticletitle{Computational game creativity}. ICCC.
\newblock


\bibitem[Lim(2017)]%
        {lim2017analysis}
\bibfield{author}{\bibinfo{person}{Francis~Pol Lim}.} \bibinfo{year}{2017}\natexlab{}.
\newblock \showarticletitle{An analysis of synchronous and asynchronous communication tools in e-learning}.
\newblock \bibinfo{journal}{\emph{Advanced Science and Technology Letters}} \bibinfo{volume}{143}, \bibinfo{number}{46} (\bibinfo{year}{2017}), \bibinfo{pages}{230--234}.
\newblock


\bibitem[Lim et~al\mbox{.}(2024)]%
        {lim2024artificial}
\bibfield{author}{\bibinfo{person}{Sue Lim}, \bibinfo{person}{Ralf Schm{\"a}lzle}, {and} \bibinfo{person}{Gary Bente}.} \bibinfo{year}{2024}\natexlab{}.
\newblock \showarticletitle{Artificial social influence via human-embodied AI agent interaction in immersive virtual reality (VR): Effects of similarity-matching during health conversations}.
\newblock \bibinfo{journal}{\emph{arXiv preprint arXiv:2406.05486}} (\bibinfo{year}{2024}).
\newblock


\bibitem[Lin et~al\mbox{.}(2022)]%
        {lin2022creative}
\bibfield{author}{\bibinfo{person}{Zhiyu Lin}, \bibinfo{person}{Rohan Agarwal}, {and} \bibinfo{person}{Mark Riedl}.} \bibinfo{year}{2022}\natexlab{}.
\newblock \showarticletitle{Creative wand: a system to study effects of communications in co-creative settings}. In \bibinfo{booktitle}{\emph{Proceedings of the AAAI Conference on Artificial Intelligence and Interactive Digital Entertainment}}, Vol.~\bibinfo{volume}{18}. \bibinfo{pages}{45--52}.
\newblock


\bibitem[Liu and Sra(2024)]%
        {liu2024dancegen}
\bibfield{author}{\bibinfo{person}{Yimeng Liu} {and} \bibinfo{person}{Misha Sra}.} \bibinfo{year}{2024}\natexlab{}.
\newblock \showarticletitle{DanceGen: Supporting Choreography Ideation and Prototyping with Generative AI}. In \bibinfo{booktitle}{\emph{Proceedings of the 2024 ACM Designing Interactive Systems Conference}}. \bibinfo{pages}{920--938}.
\newblock


\bibitem[Louie et~al\mbox{.}(2020)]%
        {louie2020novice}
\bibfield{author}{\bibinfo{person}{Ryan Louie}, \bibinfo{person}{Andy Coenen}, \bibinfo{person}{Cheng~Zhi Huang}, \bibinfo{person}{Michael Terry}, {and} \bibinfo{person}{Carrie~J. Cai}.} \bibinfo{year}{2020}\natexlab{}.
\newblock \showarticletitle{Novice-AI Music Co-Creation via AI-Steering Tools for Deep Generative Models}. In \bibinfo{booktitle}{\emph{Proceedings of the 2020 CHI Conference on Human Factors in Computing Systems}} (Honolulu, HI, USA) \emph{(\bibinfo{series}{CHI '20})}. \bibinfo{publisher}{Association for Computing Machinery}, \bibinfo{address}{New York, NY, USA}, \bibinfo{pages}{1–13}.
\newblock
\showISBNx{9781450367080}
\urldef\tempurl%
\url{https://doi.org/10.1145/3313831.3376739}
\showDOI{\tempurl}


\bibitem[Lucero et~al\mbox{.}(2016)]%
        {lucero2016designing}
\bibfield{author}{\bibinfo{person}{Andr{\'e}s Lucero}, \bibinfo{person}{Peter Dalsgaard}, \bibinfo{person}{Kim Halskov}, {and} \bibinfo{person}{Jacob Buur}.} \bibinfo{year}{2016}\natexlab{}.
\newblock \showarticletitle{Designing with cards}.
\newblock \bibinfo{journal}{\emph{Collaboration in creative design: Methods and tools}} (\bibinfo{year}{2016}), \bibinfo{pages}{75--95}.
\newblock


\bibitem[Luria and Candy(2022)]%
        {luria2022letters}
\bibfield{author}{\bibinfo{person}{Michal Luria} {and} \bibinfo{person}{Stuart Candy}.} \bibinfo{year}{2022}\natexlab{}.
\newblock \showarticletitle{Letters from the Future: Exploring Ethical Dilemmas in the Design of Social Agents}. In \bibinfo{booktitle}{\emph{Proceedings of the 2022 CHI Conference on Human Factors in Computing Systems}}. \bibinfo{pages}{1--13}.
\newblock


\bibitem[Mairesse et~al\mbox{.}(2007)]%
        {mairesse2007using}
\bibfield{author}{\bibinfo{person}{Fran{\c{c}}ois Mairesse}, \bibinfo{person}{Marilyn~A Walker}, \bibinfo{person}{Matthias~R Mehl}, {and} \bibinfo{person}{Roger~K Moore}.} \bibinfo{year}{2007}\natexlab{}.
\newblock \showarticletitle{Using linguistic cues for the automatic recognition of personality in conversation and text}.
\newblock \bibinfo{journal}{\emph{Journal of artificial intelligence research}}  \bibinfo{volume}{30} (\bibinfo{year}{2007}), \bibinfo{pages}{457--500}.
\newblock


\bibitem[Mallick et~al\mbox{.}(2024)]%
        {mallick2024pursuit}
\bibfield{author}{\bibinfo{person}{Rohit Mallick}, \bibinfo{person}{Christopher Flathmann}, \bibinfo{person}{Caitlin Lancaster}, \bibinfo{person}{Allyson Hauptman}, \bibinfo{person}{Nathan McNeese}, {and} \bibinfo{person}{Guo Freeman}.} \bibinfo{year}{2024}\natexlab{}.
\newblock \showarticletitle{The pursuit of happiness: the power and influence of AI teammate emotion in human-AI teamwork}.
\newblock \bibinfo{journal}{\emph{Behaviour \& Information Technology}} \bibinfo{volume}{43}, \bibinfo{number}{14} (\bibinfo{year}{2024}), \bibinfo{pages}{3436--3460}.
\newblock


\bibitem[Mamykina et~al\mbox{.}(2002)]%
        {mamykina2002collaborative}
\bibfield{author}{\bibinfo{person}{Lena Mamykina}, \bibinfo{person}{Linda Candy}, {and} \bibinfo{person}{Ernest Edmonds}.} \bibinfo{year}{2002}\natexlab{}.
\newblock \showarticletitle{Collaborative creativity}.
\newblock \bibinfo{journal}{\emph{Commun. ACM}} \bibinfo{volume}{45}, \bibinfo{number}{10} (\bibinfo{date}{oct} \bibinfo{year}{2002}), \bibinfo{pages}{96–99}.
\newblock
\showISSN{0001-0782}
\urldef\tempurl%
\url{https://doi.org/10.1145/570907.570940}
\showDOI{\tempurl}


\bibitem[Mascarenhas et~al\mbox{.}(2009)]%
        {mascarenhas2009cultural}
\bibfield{author}{\bibinfo{person}{Samuel Mascarenhas}, \bibinfo{person}{Jo\~{a}o Dias}, \bibinfo{person}{Nuno Afonso}, \bibinfo{person}{Sibylle Enz}, {and} \bibinfo{person}{Ana Paiva}.} \bibinfo{year}{2009}\natexlab{}.
\newblock \showarticletitle{Using rituals to express cultural differences in synthetic characters}. In \bibinfo{booktitle}{\emph{Proceedings of The 8th International Conference on Autonomous Agents and Multiagent Systems - Volume 1}} (Budapest, Hungary) \emph{(\bibinfo{series}{AAMAS '09})}. \bibinfo{publisher}{International Foundation for Autonomous Agents and Multiagent Systems}, \bibinfo{address}{Richland, SC}, \bibinfo{pages}{305–312}.
\newblock
\showISBNx{9780981738161}


\bibitem[McCormack et~al\mbox{.}(2019)]%
        {mccormack_silent_way}
\bibfield{author}{\bibinfo{person}{Jon McCormack}, \bibinfo{person}{Toby Gifford}, \bibinfo{person}{Patrick Hutchings}, \bibinfo{person}{Maria~Teresa Llano~Rodriguez}, \bibinfo{person}{Matthew Yee-King}, {and} \bibinfo{person}{Mark d'Inverno}.} \bibinfo{year}{2019}\natexlab{}.
\newblock \showarticletitle{In a Silent Way: Communication Between AI and Improvising Musicians Beyond Sound}. In \bibinfo{booktitle}{\emph{Proceedings of the 2019 CHI Conference on Human Factors in Computing Systems}} (Glasgow, Scotland Uk) \emph{(\bibinfo{series}{CHI '19})}. \bibinfo{publisher}{Association for Computing Machinery}, \bibinfo{address}{New York, NY, USA}, \bibinfo{pages}{1–11}.
\newblock
\showISBNx{9781450359702}
\urldef\tempurl%
\url{https://doi.org/10.1145/3290605.3300268}
\showDOI{\tempurl}


\bibitem[McMillan and Hwang(2002)]%
        {mcmillan2002measures}
\bibfield{author}{\bibinfo{person}{Sally~J McMillan} {and} \bibinfo{person}{Jang-Sun Hwang}.} \bibinfo{year}{2002}\natexlab{}.
\newblock \showarticletitle{Measures of perceived interactivity: An exploration of the role of direction of communication, user control, and time in shaping perceptions of interactivity}.
\newblock \bibinfo{journal}{\emph{Journal of advertising}} \bibinfo{volume}{31}, \bibinfo{number}{3} (\bibinfo{year}{2002}), \bibinfo{pages}{29--42}.
\newblock


\bibitem[Mehrotra et~al\mbox{.}(2024)]%
        {mehrotra2024systematic}
\bibfield{author}{\bibinfo{person}{Siddharth Mehrotra}, \bibinfo{person}{Chadha Degachi}, \bibinfo{person}{Oleksandra Vereschak}, \bibinfo{person}{Catholijn~M Jonker}, {and} \bibinfo{person}{Myrthe~L Tielman}.} \bibinfo{year}{2024}\natexlab{}.
\newblock \showarticletitle{A systematic review on fostering appropriate trust in Human-AI interaction: Trends, opportunities and challenges}.
\newblock \bibinfo{journal}{\emph{ACM Journal on Responsible Computing}} \bibinfo{volume}{1}, \bibinfo{number}{4} (\bibinfo{year}{2024}), \bibinfo{pages}{1--45}.
\newblock


\bibitem[Meurisch et~al\mbox{.}(2020)]%
        {meurisch2020exploring}
\bibfield{author}{\bibinfo{person}{Christian Meurisch}, \bibinfo{person}{Cristina~A Mihale-Wilson}, \bibinfo{person}{Adrian Hawlitschek}, \bibinfo{person}{Florian Giger}, \bibinfo{person}{Florian M{\"u}ller}, \bibinfo{person}{Oliver Hinz}, {and} \bibinfo{person}{Max M{\"u}hlh{\"a}user}.} \bibinfo{year}{2020}\natexlab{}.
\newblock \showarticletitle{Exploring user expectations of proactive AI systems}.
\newblock \bibinfo{journal}{\emph{Proceedings of the ACM on Interactive, Mobile, Wearable and Ubiquitous Technologies}} \bibinfo{volume}{4}, \bibinfo{number}{4} (\bibinfo{year}{2020}), \bibinfo{pages}{1--22}.
\newblock


\bibitem[Moruzzi and Margarido(2024)]%
        {moruzzi2024user}
\bibfield{author}{\bibinfo{person}{Caterina Moruzzi} {and} \bibinfo{person}{Solange Margarido}.} \bibinfo{year}{2024}\natexlab{}.
\newblock \showarticletitle{A user-centered framework for human-ai co-creativity}. In \bibinfo{booktitle}{\emph{Extended Abstracts of the CHI Conference on Human Factors in Computing Systems}}. \bibinfo{pages}{1--9}.
\newblock


\bibitem[Munnukka et~al\mbox{.}(2022)]%
        {munnukka_2022}
\bibfield{author}{\bibinfo{person}{Juha Munnukka}, \bibinfo{person}{Karoliina Talvitie-Lamberg}, {and} \bibinfo{person}{Devdeep Maity}.} \bibinfo{year}{2022}\natexlab{}.
\newblock \showarticletitle{Anthropomorphism and social presence in Human–Virtual service assistant interactions: The role of dialog length and attitudes}.
\newblock \bibinfo{journal}{\emph{Comput. Hum. Behav.}} \bibinfo{volume}{135}, \bibinfo{number}{C} (\bibinfo{date}{Oct.} \bibinfo{year}{2022}), \bibinfo{numpages}{12}~pages.
\newblock
\showISSN{0747-5632}
\urldef\tempurl%
\url{https://doi.org/10.1016/j.chb.2022.107343}
\showDOI{\tempurl}


\bibitem[Nass et~al\mbox{.}(1994)]%
        {nass1994computers}
\bibfield{author}{\bibinfo{person}{Clifford Nass}, \bibinfo{person}{Jonathan Steuer}, {and} \bibinfo{person}{Ellen~R Tauber}.} \bibinfo{year}{1994}\natexlab{}.
\newblock \showarticletitle{Computers are social actors}. In \bibinfo{booktitle}{\emph{Proceedings of the SIGCHI conference on Human factors in computing systems}}. \bibinfo{pages}{72--78}.
\newblock


\bibitem[Nass and Brave(2005)]%
        {nass2005wired}
\bibfield{author}{\bibinfo{person}{Clifford~Ivar Nass} {and} \bibinfo{person}{Scott Brave}.} \bibinfo{year}{2005}\natexlab{}.
\newblock \bibinfo{booktitle}{\emph{Wired for speech: How voice activates and advances the human-computer relationship}}.
\newblock \bibinfo{publisher}{MIT press Cambridge}.
\newblock


\bibitem[Nguyen et~al\mbox{.}(2018)]%
        {nguyen2018believe}
\bibfield{author}{\bibinfo{person}{An~T Nguyen}, \bibinfo{person}{Aditya Kharosekar}, \bibinfo{person}{Saumyaa Krishnan}, \bibinfo{person}{Siddhesh Krishnan}, \bibinfo{person}{Elizabeth Tate}, \bibinfo{person}{Byron~C Wallace}, {and} \bibinfo{person}{Matthew Lease}.} \bibinfo{year}{2018}\natexlab{}.
\newblock \showarticletitle{Believe it or not: Designing a human-ai partnership for mixed-initiative fact-checking}. In \bibinfo{booktitle}{\emph{Proceedings of the 31st Annual ACM Symposium on User Interface Software and Technology}}. \bibinfo{pages}{189--199}.
\newblock


\bibitem[Oh et~al\mbox{.}(2018)]%
        {oh2018lead}
\bibfield{author}{\bibinfo{person}{Changhoon Oh}, \bibinfo{person}{Jungwoo Song}, \bibinfo{person}{Jinhan Choi}, \bibinfo{person}{Seonghyeon Kim}, \bibinfo{person}{Sungwoo Lee}, {and} \bibinfo{person}{Bongwon Suh}.} \bibinfo{year}{2018}\natexlab{}.
\newblock \showarticletitle{I Lead, You Help but Only with Enough Details: Understanding User Experience of Co-Creation with Artificial Intelligence}. In \bibinfo{booktitle}{\emph{Proceedings of the 2018 CHI Conference on Human Factors in Computing Systems}} (Montreal QC, Canada) \emph{(\bibinfo{series}{CHI '18})}. \bibinfo{publisher}{Association for Computing Machinery}, \bibinfo{address}{New York, NY, USA}, \bibinfo{pages}{1–13}.
\newblock
\showISBNx{9781450356206}
\urldef\tempurl%
\url{https://doi.org/10.1145/3173574.3174223}
\showDOI{\tempurl}


\bibitem[Pang et~al\mbox{.}(2013)]%
        {pang2013technology}
\bibfield{author}{\bibinfo{person}{Carolyn~E Pang}, \bibinfo{person}{Carman Neustaedter}, \bibinfo{person}{Bernhard~E Riecke}, \bibinfo{person}{Erick Oduor}, {and} \bibinfo{person}{Serena Hillman}.} \bibinfo{year}{2013}\natexlab{}.
\newblock \showarticletitle{Technology preferences and routines for sharing health information during the treatment of a chronic illness}. In \bibinfo{booktitle}{\emph{Proceedings of the sigchi conference on human factors in computing systems}}. \bibinfo{pages}{1759--1768}.
\newblock


\bibitem[Park et~al\mbox{.}(2021)]%
        {park2021toddler}
\bibfield{author}{\bibinfo{person}{Junseok Park}, \bibinfo{person}{Kwanyoung Park}, \bibinfo{person}{Hyunseok Oh}, \bibinfo{person}{Ganghun Lee}, \bibinfo{person}{Minsu Lee}, \bibinfo{person}{Youngki Lee}, {and} \bibinfo{person}{Byoung-Tak Zhang}.} \bibinfo{year}{2021}\natexlab{}.
\newblock \showarticletitle{Toddler-Guidance Learning: Impacts of Critical Period on Multimodal AI Agents}. In \bibinfo{booktitle}{\emph{Proceedings of the 2021 International Conference on Multimodal Interaction}}. \bibinfo{pages}{212--220}.
\newblock


\bibitem[Poeller et~al\mbox{.}(2023)]%
        {poeller2023suspecting}
\bibfield{author}{\bibinfo{person}{Susanne Poeller}, \bibinfo{person}{Martin~Johannes Dechant}, \bibinfo{person}{Madison Klarkowski}, {and} \bibinfo{person}{Regan~L Mandryk}.} \bibinfo{year}{2023}\natexlab{}.
\newblock \showarticletitle{Suspecting sarcasm: how league of legends players dismiss positive communication in toxic environments}.
\newblock \bibinfo{journal}{\emph{Proceedings of the ACM on Human-Computer Interaction}} \bibinfo{volume}{7}, \bibinfo{number}{CHI PLAY} (\bibinfo{year}{2023}), \bibinfo{pages}{1--26}.
\newblock


\bibitem[Price et~al\mbox{.}(2022)]%
        {price_touch_2022}
\bibfield{author}{\bibinfo{person}{Sara Price}, \bibinfo{person}{Nadia Bianchi-Berthouze}, \bibinfo{person}{Carey Jewitt}, \bibinfo{person}{Nikoleta Yiannoutsou}, \bibinfo{person}{Katerina Fotopoulou}, \bibinfo{person}{Svetlana Dajic}, \bibinfo{person}{Juspreet Virdee}, \bibinfo{person}{Yixin Zhao}, \bibinfo{person}{Douglas Atkinson}, {and} \bibinfo{person}{Frederik Brudy}.} \bibinfo{year}{2022}\natexlab{}.
\newblock \showarticletitle{The Making of Meaning through Dyadic Haptic Affective Touch}.
\newblock \bibinfo{journal}{\emph{ACM Trans. Comput.-Hum. Interact.}} \bibinfo{volume}{29}, \bibinfo{number}{3}, Article \bibinfo{articleno}{21} (\bibinfo{date}{Jan.} \bibinfo{year}{2022}), \bibinfo{numpages}{42}~pages.
\newblock
\showISSN{1073-0516}
\urldef\tempurl%
\url{https://doi.org/10.1145/3490494}
\showDOI{\tempurl}


\bibitem[Razavi et~al\mbox{.}(2022)]%
        {razavi2022discourse}
\bibfield{author}{\bibinfo{person}{S~Zahra Razavi}, \bibinfo{person}{Lenhart~K Schubert}, \bibinfo{person}{Kimberly Van~Orden}, \bibinfo{person}{Mohammad~Rafayet Ali}, \bibinfo{person}{Benjamin Kane}, {and} \bibinfo{person}{Ehsan Hoque}.} \bibinfo{year}{2022}\natexlab{}.
\newblock \showarticletitle{Discourse behavior of older adults interacting with a dialogue agent competent in multiple topics}.
\newblock \bibinfo{journal}{\emph{ACM Transactions on Interactive Intelligent Systems (TiiS)}} \bibinfo{volume}{12}, \bibinfo{number}{2} (\bibinfo{year}{2022}), \bibinfo{pages}{1--21}.
\newblock


\bibitem[Rezwana and Maher(2021)]%
        {rezwanacofi}
\bibfield{author}{\bibinfo{person}{Jeba Rezwana} {and} \bibinfo{person}{Mary~Lou Maher}.} \bibinfo{year}{2021}\natexlab{}.
\newblock \showarticletitle{COFI: A Framework for Modeling Interaction in Human-AI Co-Creative Systems}.
\newblock  (\bibinfo{year}{2021}).
\newblock


\bibitem[Rezwana and Maher(2022a)]%
        {rezwana2022designing}
\bibfield{author}{\bibinfo{person}{Jeba Rezwana} {and} \bibinfo{person}{Mary~Lou Maher}.} \bibinfo{year}{2022}\natexlab{a}.
\newblock \showarticletitle{Designing Creative AI Partners with COFI: A Framework for Modeling Interaction in Human-AI Co-Creative Systems}.
\newblock \bibinfo{journal}{\emph{ACM Transactions on Computer-Human Interaction}} (\bibinfo{year}{2022}).
\newblock


\bibitem[Rezwana and Maher(2022b)]%
        {rezwana2022understanding}
\bibfield{author}{\bibinfo{person}{Jeba Rezwana} {and} \bibinfo{person}{Mary~Lou Maher}.} \bibinfo{year}{2022}\natexlab{b}.
\newblock \showarticletitle{Understanding User Perceptions, Collaborative Experience and User Engagement in Different Human-AI Interaction Designs for Co-Creative Systems}. In \bibinfo{booktitle}{\emph{Creativity and Cognition}}. \bibinfo{pages}{38--48}.
\newblock


\bibitem[Rodrigues et~al\mbox{.}(2022)]%
        {rodrigues_emoji_2022}
\bibfield{author}{\bibinfo{person}{David~L. Rodrigues}, \bibinfo{person}{Bernardo~P. Cavalheiro}, {and} \bibinfo{person}{Mar\'{\i}lia Prada}.} \bibinfo{year}{2022}\natexlab{}.
\newblock \showarticletitle{Emoji as Icebreakers? Emoji can signal distinct intentions in first time online interactions}.
\newblock \bibinfo{journal}{\emph{Telemat. Inf.}} \bibinfo{volume}{69}, \bibinfo{number}{C} (\bibinfo{date}{April} \bibinfo{year}{2022}), \bibinfo{numpages}{10}~pages.
\newblock
\showISSN{0736-5853}
\urldef\tempurl%
\url{https://doi.org/10.1016/j.tele.2022.101783}
\showDOI{\tempurl}


\bibitem[Samek et~al\mbox{.}(2023)]%
        {samek2023cosy}
\bibfield{author}{\bibinfo{person}{Fabian Samek}, \bibinfo{person}{Mathias Eulers}, \bibinfo{person}{Markus Dresel}, \bibinfo{person}{Nicole Jochems}, \bibinfo{person}{Andreas Schrader}, {and} \bibinfo{person}{Alfred Mertins}.} \bibinfo{year}{2023}\natexlab{}.
\newblock \showarticletitle{CoSy-AI enhanced assistance system for face to face communication trainings in higher healthcare education: AI enhanced assistance system for face to face communication trainings in higher healthcare education}. In \bibinfo{booktitle}{\emph{Proceedings of the 16th International Conference on PErvasive Technologies Related to Assistive Environments}}. \bibinfo{pages}{457--460}.
\newblock


\bibitem[Sicilia et~al\mbox{.}(2024)]%
        {sicilia2024humbel}
\bibfield{author}{\bibinfo{person}{Anthony Sicilia}, \bibinfo{person}{Jennifer Gates}, {and} \bibinfo{person}{Malihe Alikhani}.} \bibinfo{year}{2024}\natexlab{}.
\newblock \showarticletitle{HumBEL: A Human-in-the-Loop Approach for Evaluating Demographic Factors of Language Models in Human-Machine Conversations}. In \bibinfo{booktitle}{\emph{Proceedings of the 18th Conference of the European Chapter of the Association for Computational Linguistics (Volume 1: Long Papers)}}. \bibinfo{pages}{1127--1143}.
\newblock


\bibitem[Sonlu et~al\mbox{.}(2021)]%
        {sonlu2021conversational}
\bibfield{author}{\bibinfo{person}{Sinan Sonlu}, \bibinfo{person}{U{\u{g}}ur G{\"u}d{\"u}kbay}, {and} \bibinfo{person}{Funda Durupinar}.} \bibinfo{year}{2021}\natexlab{}.
\newblock \showarticletitle{A conversational agent framework with multi-modal personality expression}.
\newblock \bibinfo{journal}{\emph{ACM Transactions on Graphics (TOG)}} \bibinfo{volume}{40}, \bibinfo{number}{1} (\bibinfo{year}{2021}), \bibinfo{pages}{1--16}.
\newblock


\bibitem[Striner et~al\mbox{.}(2022)]%
        {striner2022co}
\bibfield{author}{\bibinfo{person}{Alina Striner}, \bibinfo{person}{Thomas R{\"o}ggla}, \bibinfo{person}{Mikel Zorrilla}, \bibinfo{person}{Sergio Cabrero~Barros}, \bibinfo{person}{Stefano Masneri}, \bibinfo{person}{H{\'e}ctor Rivas~Pagador}, \bibinfo{person}{Irene Calvis}, \bibinfo{person}{Jie Li}, {and} \bibinfo{person}{Pablo Cesar}.} \bibinfo{year}{2022}\natexlab{}.
\newblock \showarticletitle{The Co-Creation Space: Supporting Asynchronous Artistic Co-creation Dynamics}. In \bibinfo{booktitle}{\emph{Companion Publication of the 2022 Conference on Computer Supported Cooperative Work and Social Computing}}. \bibinfo{pages}{18--22}.
\newblock


\bibitem[van Berkel et~al\mbox{.}(2021)]%
        {van2021proactive}
\bibfield{author}{\bibinfo{person}{Niels van Berkel}, \bibinfo{person}{Mikael~B. Skov}, {and} \bibinfo{person}{Jesper Kjeldskov}.} \bibinfo{year}{2021}\natexlab{}.
\newblock \showarticletitle{Human-AI interaction: intermittent, continuous, and proactive}.
\newblock \bibinfo{journal}{\emph{Interactions}} \bibinfo{volume}{28}, \bibinfo{number}{6} (\bibinfo{date}{nov} \bibinfo{year}{2021}), \bibinfo{pages}{67–71}.
\newblock
\showISSN{1072-5520}
\urldef\tempurl%
\url{https://doi.org/10.1145/3486941}
\showDOI{\tempurl}


\bibitem[Vodrahalli et~al\mbox{.}(2022)]%
        {vodrahalli2022humans}
\bibfield{author}{\bibinfo{person}{Kailas Vodrahalli}, \bibinfo{person}{Roxana Daneshjou}, \bibinfo{person}{Tobias Gerstenberg}, {and} \bibinfo{person}{James Zou}.} \bibinfo{year}{2022}\natexlab{}.
\newblock \showarticletitle{Do humans trust advice more if it comes from ai? an analysis of human-ai interactions}. In \bibinfo{booktitle}{\emph{Proceedings of the 2022 AAAI/ACM Conference on AI, Ethics, and Society}}. \bibinfo{pages}{763--777}.
\newblock


\bibitem[Vossen et~al\mbox{.}(2009)]%
        {vossen2009social}
\bibfield{author}{\bibinfo{person}{Suzanne Vossen}, \bibinfo{person}{Jaap Ham}, {and} \bibinfo{person}{Cees Midden}.} \bibinfo{year}{2009}\natexlab{}.
\newblock \showarticletitle{Social influence of a persuasive agent: the role of agent embodiment and evaluative feedback}. In \bibinfo{booktitle}{\emph{Proceedings of the 4th International Conference on Persuasive Technology}}. \bibinfo{pages}{1--7}.
\newblock


\bibitem[V{\"o}ssing et~al\mbox{.}(2022)]%
        {vossing2022designing}
\bibfield{author}{\bibinfo{person}{Michael V{\"o}ssing}, \bibinfo{person}{Niklas K{\"u}hl}, \bibinfo{person}{Matteo Lind}, {and} \bibinfo{person}{Gerhard Satzger}.} \bibinfo{year}{2022}\natexlab{}.
\newblock \showarticletitle{Designing transparency for effective human-AI collaboration}.
\newblock \bibinfo{journal}{\emph{Information Systems Frontiers}} \bibinfo{volume}{24}, \bibinfo{number}{3} (\bibinfo{year}{2022}), \bibinfo{pages}{877--895}.
\newblock


\bibitem[Wang et~al\mbox{.}(2023)]%
        {wang_haptic_2023}
\bibfield{author}{\bibinfo{person}{Yiwen Wang}, \bibinfo{person}{Ziming Li}, \bibinfo{person}{Pratheep~Kumar Chelladurai}, \bibinfo{person}{Wendy Dannels}, \bibinfo{person}{Tae Oh}, {and} \bibinfo{person}{Roshan~L Peiris}.} \bibinfo{year}{2023}\natexlab{}.
\newblock \showarticletitle{Haptic-Captioning: Using Audio-Haptic Interfaces to Enhance Speaker Indication in Real-Time Captions for Deaf and Hard-of-Hearing Viewers}. In \bibinfo{booktitle}{\emph{Proceedings of the 2023 CHI Conference on Human Factors in Computing Systems}} (Hamburg, Germany) \emph{(\bibinfo{series}{CHI '23})}. \bibinfo{publisher}{Association for Computing Machinery}, \bibinfo{address}{New York, NY, USA}, Article \bibinfo{articleno}{781}, \bibinfo{numpages}{14}~pages.
\newblock
\showISBNx{9781450394215}
\urldef\tempurl%
\url{https://doi.org/10.1145/3544548.3581076}
\showDOI{\tempurl}


\bibitem[Wegner(1997)]%
        {wegner1997interaction}
\bibfield{author}{\bibinfo{person}{Peter Wegner}.} \bibinfo{year}{1997}\natexlab{}.
\newblock \showarticletitle{Why interaction is more powerful than algorithms}.
\newblock \bibinfo{journal}{\emph{Commun. ACM}} \bibinfo{volume}{40}, \bibinfo{number}{5} (\bibinfo{year}{1997}), \bibinfo{pages}{80--91}.
\newblock


\bibitem[Weld and Bansal(2018)]%
        {weld2018intelligible}
\bibfield{author}{\bibinfo{person}{Daniel~S Weld} {and} \bibinfo{person}{Gagan Bansal}.} \bibinfo{year}{2018}\natexlab{}.
\newblock \showarticletitle{Intelligible artificial intelligence}.
\newblock \bibinfo{journal}{\emph{ArXiv e-prints, March 2018}} (\bibinfo{year}{2018}).
\newblock


\bibitem[W{\"o}lfel and Merritt(2013)]%
        {wolfel2013method}
\bibfield{author}{\bibinfo{person}{Christiane W{\"o}lfel} {and} \bibinfo{person}{Timothy Merritt}.} \bibinfo{year}{2013}\natexlab{}.
\newblock \showarticletitle{Method card design dimensions: A survey of card-based design tools}. In \bibinfo{booktitle}{\emph{Human-Computer Interaction--INTERACT 2013: 14th IFIP TC 13 International Conference, Cape Town, South Africa, September 2-6, 2013, Proceedings, Part I 14}}. Springer, \bibinfo{pages}{479--486}.
\newblock


\bibitem[Wu et~al\mbox{.}(2021)]%
        {wu2021ai}
\bibfield{author}{\bibinfo{person}{Zhuohao Wu}, \bibinfo{person}{Danwen Ji}, \bibinfo{person}{Kaiwen Yu}, \bibinfo{person}{Xianxu Zeng}, \bibinfo{person}{Dingming Wu}, {and} \bibinfo{person}{Mohammad Shidujaman}.} \bibinfo{year}{2021}\natexlab{}.
\newblock \showarticletitle{AI Creativity and the Human-AI Co-creation Model}. In \bibinfo{booktitle}{\emph{International Conference on Human-Computer Interaction}}. \bibinfo{pages}{171--190}.
\newblock


\bibitem[Xia and Chen(2023)]%
        {xia2023review}
\bibfield{author}{\bibinfo{person}{Yan Xia} {and} \bibinfo{person}{Yue Chen}.} \bibinfo{year}{2023}\natexlab{}.
\newblock \showarticletitle{A Review of How Team Creativity is Affected by the Design of Communication Tools}. In \bibinfo{booktitle}{\emph{International Conference on Human-Computer Interaction}}. Springer, \bibinfo{pages}{297--314}.
\newblock


\bibitem[Yang et~al\mbox{.}(2020)]%
        {yang2020re}
\bibfield{author}{\bibinfo{person}{Qian Yang}, \bibinfo{person}{Aaron Steinfeld}, \bibinfo{person}{Carolyn Ros{\'e}}, {and} \bibinfo{person}{John Zimmerman}.} \bibinfo{year}{2020}\natexlab{}.
\newblock \showarticletitle{Re-examining whether, why, and how human-AI interaction is uniquely difficult to design}. In \bibinfo{booktitle}{\emph{Proceedings of the 2020 chi conference on human factors in computing systems}}. \bibinfo{pages}{1--13}.
\newblock


\bibitem[Yeh et~al\mbox{.}(2022)]%
        {yeh2022guide}
\bibfield{author}{\bibinfo{person}{Su-Fang Yeh}, \bibinfo{person}{Meng-Hsin Wu}, \bibinfo{person}{Tze-Yu Chen}, \bibinfo{person}{Yen-Chun Lin}, \bibinfo{person}{XiJing Chang}, \bibinfo{person}{You-Hsuan Chiang}, {and} \bibinfo{person}{Yung-Ju Chang}.} \bibinfo{year}{2022}\natexlab{}.
\newblock \showarticletitle{How to guide task-oriented chatbot users, and when: A mixed-methods study of combinations of chatbot guidance types and timings}. In \bibinfo{booktitle}{\emph{Proceedings of the 2022 CHI Conference on Human Factors in Computing Systems}}. \bibinfo{pages}{1--16}.
\newblock


\bibitem[Zhang et~al\mbox{.}(2023)]%
        {zhang2023investigating}
\bibfield{author}{\bibinfo{person}{Rui Zhang}, \bibinfo{person}{Wen Duan}, \bibinfo{person}{Christopher Flathmann}, \bibinfo{person}{Nathan McNeese}, \bibinfo{person}{Guo Freeman}, {and} \bibinfo{person}{Alyssa Williams}.} \bibinfo{year}{2023}\natexlab{}.
\newblock \showarticletitle{Investigating AI teammate communication strategies and their impact in human-AI teams for effective teamwork}.
\newblock \bibinfo{journal}{\emph{Proceedings of the ACM on Human-Computer Interaction}} \bibinfo{volume}{7}, \bibinfo{number}{CSCW2} (\bibinfo{year}{2023}), \bibinfo{pages}{1--31}.
\newblock


\bibitem[Zhang et~al\mbox{.}(2024)]%
        {zhang2024verbal}
\bibfield{author}{\bibinfo{person}{Rui Zhang}, \bibinfo{person}{Wen Duan}, \bibinfo{person}{Christopher Flathmann}, \bibinfo{person}{Nathan McNeese}, \bibinfo{person}{Bart Knijnenburg}, {and} \bibinfo{person}{Guo Freeman}.} \bibinfo{year}{2024}\natexlab{}.
\newblock \showarticletitle{Verbal vs. Visual: How Humans Perceive and Collaborate with AI Teammates Using Different Communication Modalities in Various Human-AI Team Compositions}.
\newblock \bibinfo{journal}{\emph{Proceedings of the ACM on Human-Computer Interaction}} \bibinfo{volume}{8}, \bibinfo{number}{CSCW2} (\bibinfo{year}{2024}), \bibinfo{pages}{1--34}.
\newblock


\bibitem[Zhang et~al\mbox{.}(2021)]%
        {zhang2021ideal}
\bibfield{author}{\bibinfo{person}{Rui Zhang}, \bibinfo{person}{Nathan~J McNeese}, \bibinfo{person}{Guo Freeman}, {and} \bibinfo{person}{Geoff Musick}.} \bibinfo{year}{2021}\natexlab{}.
\newblock \showarticletitle{" An ideal human" expectations of AI teammates in human-AI teaming}.
\newblock \bibinfo{journal}{\emph{Proceedings of the ACM on Human-Computer Interaction}} \bibinfo{volume}{4}, \bibinfo{number}{CSCW3} (\bibinfo{year}{2021}), \bibinfo{pages}{1--25}.
\newblock


\bibitem[Zhao et~al\mbox{.}(2024)]%
        {zhao2024tailoring}
\bibfield{author}{\bibinfo{person}{Xinyan Zhao}, \bibinfo{person}{Yuan Sun}, \bibinfo{person}{Wenlin Liu}, {and} \bibinfo{person}{Chau-Wai Wong}.} \bibinfo{year}{2024}\natexlab{}.
\newblock \showarticletitle{Tailoring Generative AI Chatbots for Multiethnic Communities in Disaster Preparedness Communication: Extending the CASA Paradigm}.
\newblock \bibinfo{journal}{\emph{arXiv preprint arXiv:2406.08411}} (\bibinfo{year}{2024}).
\newblock


\bibitem[Zhou and Hu(2024)]%
        {zhou2024beyond}
\bibfield{author}{\bibinfo{person}{Jijie Zhou} {and} \bibinfo{person}{Yuhan Hu}.} \bibinfo{year}{2024}\natexlab{}.
\newblock \showarticletitle{Beyond Words: Infusing Conversational Agents with Human-like Typing Behaviors}. In \bibinfo{booktitle}{\emph{Proceedings of the 6th ACM Conference on Conversational User Interfaces}}. \bibinfo{pages}{1--12}.
\newblock


\bibitem[Zhou and Gorman(2024)]%
        {zhou2024impact}
\bibfield{author}{\bibinfo{person}{Shiwen Zhou} {and} \bibinfo{person}{Jamie~C Gorman}.} \bibinfo{year}{2024}\natexlab{}.
\newblock \showarticletitle{The Impact of Communication Timing and Sequencing on Team Performance: A Comparative Study of Human-AI and All-Human Teams}. In \bibinfo{booktitle}{\emph{Proceedings of the Human Factors and Ergonomics Society Annual Meeting}}, Vol.~\bibinfo{volume}{68}. SAGE Publications Sage CA: Los Angeles, CA, \bibinfo{pages}{1769--1774}.
\newblock


\bibitem[Zhu et~al\mbox{.}(2018)]%
        {zhu2018explainable}
\bibfield{author}{\bibinfo{person}{Jichen Zhu}, \bibinfo{person}{Antonios Liapis}, \bibinfo{person}{Sebastian Risi}, \bibinfo{person}{Rafael Bidarra}, {and} \bibinfo{person}{G~Michael Youngblood}.} \bibinfo{year}{2018}\natexlab{}.
\newblock \showarticletitle{Explainable AI for designers: A human-centered perspective on mixed-initiative co-creation}. In \bibinfo{booktitle}{\emph{2018 IEEE conference on computational intelligence and games (CIG)}}. IEEE, \bibinfo{pages}{1--8}.
\newblock


\end{thebibliography}


\end{document}